\documentclass[onecolumn,12pt,journal,final]{IEEEtran} 

\def\ps@headings{%
\def\@oddhead{\mbox{}\scriptsize\rightmark \hfil \thepage}%
\def\@evenhead{\scriptsize\thepage \hfil \leftmark\mbox{}}%
\def\@oddfoot{}%
\def\@evenfoot{}}
\makeatother 

\IEEEoverridecommandlockouts
\usepackage{bbm}
\usepackage{amsfonts}
\usepackage[dvips]{graphicx}
\usepackage{times}
\usepackage{cite}
\usepackage{amsmath}
\usepackage{array}
\usepackage{amssymb}

\usepackage{stfloats}
\usepackage{graphicx}
\usepackage{footnote}
\usepackage{booktabs}
\usepackage{array}
\usepackage{algorithmic}
\usepackage{algorithm}
\usepackage{subeqnarray}
\usepackage{cases}
\usepackage{threeparttable}
\usepackage{color}
\usepackage{hyperref}
\usepackage{epstopdf}
\usepackage{cleveref}
\usepackage{bm}
\usepackage{subcaption}
\usepackage{dsfont}

\usepackage[labelformat=simple]{subcaption}

\newtheorem{proposition}{Proposition}
\newtheorem{lemma}{Lemma}

\begin{document}

\title{Joint Base Station Clustering and Beamforming for Non-Orthogonal Multicast and Unicast Transmission with Backhaul Constraints}
\author{Erkai~Chen, Meixia~Tao, and Ya-Feng~Liu \\
\thanks{This paper was presented in part at the IEEE GLOBECOM 2017 \cite{Chen_GC17}. }
\thanks{E.~Chen and M.~Tao are with the Department of Electronic Engineering at Shanghai Jiao Tong University, Shanghai 200240, China (email: cek1006@sjtu.edu.cn; mxtao@sjtu.edu.cn).}
\thanks{Y.-F.~Liu is with the State Key Laboratory of Scientific and Engineering Computing, Institute of Computational Mathematics and Scientific/Engineering Computing, Academy of Mathematics and Systems Science, Chinese Academy of Sciences, Beijing 100190, China (e-mail: yafliu@lsec.cc.ac.cn).}
}
\maketitle

\begin{abstract}
The demand for providing multicast services in cellular networks is continuously and fastly increasing. In this work, we propose a non-orthogonal transmission framework based on layered-division multiplexing (LDM) to support multicast and unicast services concurrently in cooperative multi-cell cellular networks with limited backhaul capacity. We adopt a two-layer LDM structure where the first layer is intended for multicast services, the second layer is for unicast services, and the two layers are superposed with different beamformers. Each user decodes the multicast message first, subtracts it, and then decodes its dedicated unicast message. We formulate a joint multicast and unicast beamforming problem with adaptive base station clustering that aims to maximize the weighted sum of the multicast rate and the unicast rate under per-BS power and backhaul constraints. To solve the problem, we first develop a branch-and-bound algorithm to find its global optimum. We then reformulate the problem as a sparse beamforming problem and propose a low-complexity algorithm based on convex-concave procedure. Simulation results demonstrate the significant superiority of the proposed LDM-based non-orthogonal scheme over orthogonal schemes in terms of the achievable multicast-unicast rate region.
\end{abstract}


\section{Introduction}
The broadcast nature of the wireless medium makes multicasting an efficient point-to-multipoint communication mechanism to deliver a same content concurrently to multiple interested users or devices. Recently, multicast services have been gaining increasing interests in cellular networks due to emerging applications such as live video streaming, venue casting, proactive multimedia content pushing, software updates, and public group communications \cite{Ericsson_multicast}. 
In conventional cellular networks, multicast services have been allocated different time or frequency resources from those allocated to unicast services and adopt single-frequency network (SFN) transmission, as in the 3GPP specifications known as LTE-multicast \cite{LTE_eMBMS_Rel11_CM12}. However, such orthogonal resource sharing and transmission scheme has low spectrum efficiency and can significantly degrade the performance of the existing unicast services. 
Techniques that allow cellular networks to carry multicast and unicast services jointly in a more spectrum-efficient way are highly desirable. 
There are also many practical scenarios where a user needs to receive both multicast and unicast signals at the same time. For example, the network operator would like to offer multicast services like proactive content pushing, automatic software updates, and public group announcements to its subscribers without interrupting their on-going unicast services. Content providers can also embed personalized information (e.g., preferred subtitles and targeted advertisements) via unicast transmission along the multicast-based video streaming.

\subsection{Related Works} \label{sec:Related Works}
To address the need of joint multicast and unicast transmission in cellular networks, several research efforts have been made. 
One possible way is to use MIMO spatial multiplexing where all the multicast and unicast messages are transmitted with different beamformers and each is decoded at its desired receiver by treating all other signals as noise \cite{Silva_PIMRC06_spatial_uni_multi,Larsson_Broadcasting_TWC16,content_centric_TWC16,Oskari_EE_joint_unicast_multicast_SPAWC17}. 
The authors in \cite{Silva_PIMRC06_spatial_uni_multi} studied the adaptive beamforming for the coexistence of the multicast and unicast services in a multi-user multi-carrier system.
The authors in \cite{Larsson_Broadcasting_TWC16} introduced a joint beamforming and broadcasting technique, which exploits the surplus of spatial degrees of freedom in massive MIMO systems. Its main idea is to broadcast a common message to users whose channel state information (CSI) is unavailable and to beamform unicast messages to users whose CSI is available.
The authors in \cite{content_centric_TWC16} introduced a content-centric beamforming design for content delivery in a cache-enabled radio access network. It includes the joint multicast and unicast beamforming problem as a special case when some users request the same content and others request distinct contents for each. 
The authors in \cite{Oskari_EE_joint_unicast_multicast_SPAWC17} studied energy-efficient joint transmit and receive beamforming in a multi-cell multi-user MIMO system, where the users can receive unicast messages in addition to the group-specific multicast messages at the same time. Different messages are separated in the spatial domain at the users which are equipped with multiple receive antennas.
Instead of using spatial multiplexing, another way is to adopt superposition coding to deliver both multicast and unicast services simultaneously. Each receiver decodes its desired multicast and unicast messages successively by using the successive interference cancellation (SIC)-based multi-user detection \cite{Kim_Superposition_Mag08, Sun_PIMRC10_uni_multi_two_user,Choi_TCOM15_Superposition,Clerckx_ICC25_joint_mult_bro_CSIT}.
More specifically, the scheduling and resource sharing problem for the superposition of broadcast and unicast in wireless cellular systems is studied in \cite{Kim_Superposition_Mag08}. 
The MIMO beamforming problem in a simple case with only two users (i.e., near and far) is studied in \cite{Sun_PIMRC10_uni_multi_two_user,Choi_TCOM15_Superposition}.
The performance of the joint multicast and unicast transmission with partial CSI is studied in \cite{Clerckx_ICC25_joint_mult_bro_CSIT}. 
A more general scenario is considered in \cite{Medi_TWC17_SGD} for a multi-cell network, where each base station (BS) sends multiple independent multicast messages and each user can decode an arbitrary subset of these multicast messages from all BSs using successive group decoding.

Recently, layered-division multiplexing (LDM), a form of non-orthogonal multiplexing technology \cite{Zhang_LDM_ToB16}, has been introduced in cellular networks for joint multicast and unicast transmission \cite{Simeone_LDM_GLOBECOM16,Liu_SPAWC17}. It is a key technology for next-generation terrestrial digital television standard ATSC 3.0 \cite{Simeone_LDM_ToB15}. LDM applies a layered transmission structure to transmit multiple signals with different power levels and robustness for different services and reception environment. A receiver can decode the upper layer most robust signal first, cancel it from the received signal, and then decode the next layer signal.
By using LDM, a joint beamforming design algorithm is proposed in \cite{Simeone_LDM_GLOBECOM16} for minimizing the total transmit power under constraints on the user specific unicast rate and the common multicast rate.
Note that, the work \cite{Simeone_LDM_GLOBECOM16} only considered a fixed BS clustering scheme for both multicast and unicast beamformers without taking channel dynamics into account. 
The authors in \cite{Liu_SPAWC17} considered a similar problem but introduced a group-sparse encouraging penalty in the objective function to reduce signaling overhead among different BSs.
However, neither of the above works explicitly considered backhaul constraints. In practice, each BS is usually connected to the core network and cooperates with other BSs via a backhaul link with a finite capacity. Thus, the joint transmission among multiple BSs needs to take the backhaul constraints into account explicitly.

In a different line of research on non-orthogonal multiplexing, the power-domain non-orthogonal multiple access (NOMA) \cite{Saito_NOMA_VTC13,Ding_NOMA_CM17} and the rate splitting (RS) \cite{Bruno_RS_CM16,Brono_RS_TSP16} have been studied as promising technologies to increase system performance in wireless networks. 
In power-domain NOMA, two users with different channel conditions (i.e., poor and strong) are served on the same time/frequency/code resource with different power levels. The user with strong channel condition decodes the message of the user with poor channel condition first, cancels it, and then decodes its own message. Thus, the message of the user with poor channel condition can be viewed as a common message intended to both users.
In RS, each user's message is split into a common part and a private part. All common parts are packed into one common message, which is superimposed and simultaneously transmitted with the unicast messages. It has been studied as a promising strategy for robust transmission with imperfect CSI at the transmitter \cite{Bruno_RS_CM16}.
It is worth remarking that in the power-domain NOMA with MIMO beamforming, multiple messages share a same beamforming vector but with different powers \cite{Saito_NOMA_VTC13,Ding_NOMA_CM17}. On the other hand, the LDM-based non-orthogonal transmission assigns a dedicated beamforming vector for each message \cite{Simeone_LDM_GLOBECOM16,Liu_SPAWC17}, i.e., the messages are superposed with different beamformers.
We also remark that while the RS signal model resembles the LDM-based non-orthogonal transmission, the role of the multicast message is fundamentally different. The multicast message in RS encapsulates parts of the unicast messages, and is decoded by all users for interference mitigation, although not entirely required by themselves \cite{Brono_RS_TSP16}, while the multicast message in the LDM-based non-orthogonal transmission carries common information intended as a whole for all users.

\subsection{Contributions} \label{sec:Contributions} 
In this paper, we propose a new LDM-based non-orthogonal transmission framework for multicast and unicast services in multi-cell cooperative cellular networks with backhaul constraints.
As in \cite{Simeone_LDM_GLOBECOM16,Liu_SPAWC17}, we adopt a two-layer LDM structure where the first layer is intended for multicast services and the second layer is for unicast services. The two layers are superposed with different network-wide beamformers which are potentially (group) sparse due to the backhaul constraints. Each user decodes the multicast message first, subtracts it, and then decodes its unicast message. Different from \cite{Simeone_LDM_GLOBECOM16,Liu_SPAWC17}, we consider dynamic BS clustering for each message with respect to instantaneous channel conditions and the per-BS backhaul constraints. 
Under this non-orthogonal transmission framework, we seek the maximum achievable rates of both multicast and unicast services under the peak power and peak backhaul constraints on each individual BS by the joint design of BS clustering and beamforming.

The main contributions of this paper are summarized as follows:
\begin{itemize} 
	\item 
	\emph{New Problem Formulation:} 
	We formulate a mixed-integer non-linear programming (MINLP) problem for the joint design of BS clustering and beamforming to maximize the weighted sum of the multicast rate and the unicast rate under the per-BS power and backhaul constraints. By varying the weighting parameter, we can set different priorities on the multicast and unicast services and hence obtain different achievable multicast-unicast rate pairs. 
	Note that this problem is challenging due to the combinatorial nature of the BS clustering variables and the coupling between the BS clustering variables and rate variables in the backhaul constraints.
	\item 
	\emph{Optimal Branch-and-Bound Algorithm:}
	We design a branch-and-bound (BB) algorithm to find the global optimal solution of the above formulated problem with guaranteed convergence by using the convex relaxation techniques in \cite{Liu_SPAWC17} and \cite{Biconvex83}. 
	Although with (theoretically) high computational complexity, the BB-based algorithm serves as a benchmark for evaluating the performance of other heuristic or local algorithms for the same problem.
	\item 
	\emph{High-Performance Low-Complexity Algorithm:}
	Considering the practical implementation, we also design a low-complexity algorithm. Simulation results show that it can achieve high performance that is very close to the optimum. Specifically, we first reformulate the joint design problem as an equivalent sparse beamforming problem. The equivalent problem is still challenging due to that the per-BS backhaul constraints involve not only the discontinuous $\ell_0$-norm but also the product of two non-convex functions.
	Then we use a concave smooth function to approximate the discontinuous $\ell_0$-norm and use difference of squares to rewrite the product form. By doing so, the problem is then transformed (with approximation) into a difference of convex (DC) programming problem, for which a stationary solution can be obtained efficiently by using the convex-concave procedure (CCP) with guaranteed convergence. 
	\item
	\emph{Promising Simulation Results:}
	Simulation results show that our proposed low-complexity algorithm can achieve performance that is very close to the global optimum.
	The results also demonstrate that our proposed LDM-based non-orthogonal scheme can achieve a significantly larger multicast-unicast rate region than orthogonal schemes. This indicates that our proposed LDM-based non-orthogonal transmission can serve as an efficient scheme to incorporate multicast and unicast services in cellular networks.
\end{itemize}

\subsection{Organization and Notations}
The rest of the paper is organized as follows. Section \ref{sec:System Model and Problem Formulation} introduces the system model and the problem formulation. Section \ref{sec:BB-based Optimal Algorithm} provides the details of the proposed optimal solution based on the BB method. A CCP-based low-complexity algorithm is developed in Section \ref{sec:CCP-based Low-complexity Algorithm}. Simulation results are provided in Section \ref{sec:Simulation Results}. Finally, we conclude the paper in Section \ref{sec:Conclusion}.

\emph{Notations}: 
The operators $(\cdot)^T$ and $(\cdot)^H$ correspond to the transpose and Hermitian transpose, respectively. $\mathcal{CN}(\delta,\sigma^2)$ represents a complex Gaussian distribution with mean $\delta$ and variance $\sigma^2$. The real and imaginary parts of a complex number $x$ are denoted by $\Re\{x\}$ and $\Im\{x\}$, respectively. Finally, $\mathbf{0}_L$ denotes the all-zero vector of dimension $L$.

\begin{figure}[tbp]
\begin{centering}
\includegraphics[scale=.32]{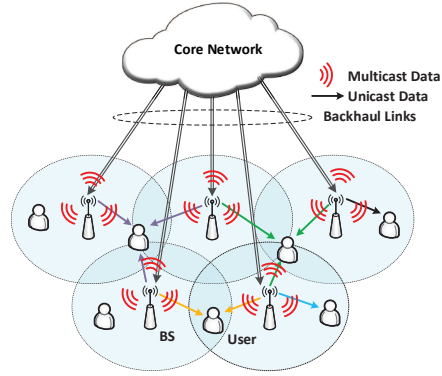}
 \caption{\small{System model of non-orthogonal multicast and unicast downlink transmission.}}\label{fig:fig_system_model}
\end{centering}
\end{figure}

\section{System Model and Problem Formulation} \label{sec:System Model and Problem Formulation}
\subsection{System Model} \label{sec:System Model}
Consider the downlink transmission of a backhaul-constrained cooperative multi-cell cellular network, where $N$ BSs, each equipped with $L$ transmit antennas, collectively provide hybrid multicast and unicast services, as shown in Fig. \ref{fig:fig_system_model}. 
In each scheduling slot, there are $K$ active users, each with a single antenna. Each user has a dedicated unicast request and subscribes to a group-specific multicast service.
In general, there can be multiple multicast groups according to different multicast service subscriptions. In this paper, for ease of the notation, we focus on one multicast group only, i.e., there is one common multicast message intended for all users. The results obtained in this paper can be easily extended to the multi-group scenario.

The backhaul link that connects each BS $n$ to the core network, which has access to all service or content providers, is subject to a peak capacity constraint of $C_n$ bits/s, for all $n \in \mathcal{N} \triangleq \{1,\dots, N\}$. Due to such backhaul constraints, not every BS can participate in the transmission of every multicast and unicast messages. 
Let the binary variable $s_{0,n} = 1$ indicate that the $n$-th BS belongs to the serving BS cluster of the multicast message and $s_{0,n} = 0$ otherwise. Similarly, let $s_{k,n} = 1$ indicate that the $n$-th BS belongs to the serving BS cluster of the unicast message for user $k$ and $s_{k,n} = 0$ otherwise. 

Let $x_{0} \in \mathbb{C}$ denote the multicast message intended for all $K$ users and $x_{k} \in \mathbb{C}$ the unicast message intended for user $k$, for all $k \in \mathcal{K} \triangleq \{1,\dots,K\}$, all with normalized power of $1$.
We adopt a two-layer LDM structure where the first layer is intended for the multicast service, the second layer is for unicast services, and the two layers are superposed with different beamformers at each BS.
Let $\mathbf{w}_{0,n} \in \mathbb{C}^{L \times 1}$ denote the beamforming vector at BS $n$ for the multicast message $x_{0}$ and $\mathbf{w}_{k,n} \in \mathbb{C}^{L \times 1}$ denote the beamforming vector at BS $n$ for the unicast message $x_{k}$, respectively.
The transmit signal of BS $n$ can be written as
\begin{align}
\mathbf{z}_{n} = \underbrace{\mathbf{w}_{0,n} x_{0}}_{\text{multicast layer}} + \underbrace{\sum_{k=1}^K \mathbf{w}_{k,n} x_{k}}_{\text{unicast layer}}.
\end{align}
The total transmit power of the multicast layer and the unicast layer on each BS $n$ is subject to a peak power constraint as
\begin{align}
\| \mathbf{w}_{0,n} \|_2^2 + \sum_{k=1}^K \| \mathbf{w}_{k,n} \|_2^2 \leq P_n, \label{cons:power-constraint}
\end{align}
where $P_n$ is the peak transmit power of the $n$-th BS. Note that $\| \mathbf{w}_{0,n} \|_2^2 = 0$ ($\| \mathbf{w}_{k,n} \|_2^2 = 0$) if $s_{0,n} = 0$ ($s_{k,n} = 0$), which implies that BS $n$ does not participate in the transmission of message $x_{0}$ ($x_{k}$). Thus, we have the following constraint:
\begin{align}
\lVert \mathbf{w}_{k,n} \rVert_2^2 \leq s_{k,n} P_n, ~\forall~ k \in \widehat{\mathcal{K}},~n \in \mathcal{N},
\end{align}
where $\widehat{\mathcal{K}} \triangleq \mathcal{K} \cup \{0\}$ is the index set of all $K$ unicast messages and one multicast message. 

The received signal at the $k$-th user is expressed as
\begin{align}
y_{k}
&= \sum_{n = 1}^N \mathbf{h}_{k,n}^H \mathbf{z}_{n} + n_k, \nonumber \\
&= \sum_{n = 1}^N \mathbf{h}_{k,n}^H \mathbf{w}_{0,n} x_{0} + \sum_{n = 1}^N \sum_{j=1}^K \mathbf{h}_{k,n}^H \mathbf{w}_{j,n} x_{j} + n_k, \nonumber \\
&= \underbrace{\mathbf{h}_{k}^H \mathbf{w}_{0} x_{0}}_{\text{multicast signal}} + \underbrace{\mathbf{h}_{k}^H \mathbf{w}_{k} x_{k}}_{\text{unicast signal}} + \underbrace{\sum_{j=1, \, j \neq k}^K \mathbf{h}_{k}^H \mathbf{w}_{j} x_{j}}_{\text{inter-user interference}} + \underbrace{n_k}_{\text{noise}},
\end{align}
where $\mathbf{h}_k = [\mathbf{h}_{k,1}^H,\mathbf{h}_{k,2}^H, \dots, \mathbf{h}_{k,N}^H ]^H \in \mathbb{C}^{NL\times 1}$ is the network-wide channel vector between all BSs and user $k$, $\mathbf{w}_{0} \in \mathbb{C}^{NL\times 1}$ and $\mathbf{w}_{k} \in \mathbb{C}^{NL\times 1}$ are the network-wide beamforming vectors defined in a similar manner, and $n_k \sim \mathcal{CN}(0,\sigma_k^2)$ is the additive white Gaussian noise at user $k$. Without loss of generality, we assume that all of the channel vectors are linearly independent. We also assume that perfect CSI is available at the core network for joint processing and all BSs can precisely synchronize with each other, and focus on the beamforming design to evaluate the advantages of the proposed non-orthogonal multicast and unicast transmission framework. Typically, CSI can be collected by estimating it at each user and feeding it back to the BS via a feedback channel in frequency-division-duplex (FDD) systems, or through uplink channel estimation in time-division-duplex (TDD) systems. Each BS collects its own CSI and sends it to the central controller in the core network via its backhaul link. For time synchronization, a combination of global positioning system (GPS) and network synchronization protocol can be used for synchronizing the primary clock as well as the frame structure in distant BSs \cite{SWCS08_timesynchronization}.

At each receiver, SIC is used to decode the multicast message and the desired unicast message successively while treating the unicast signals of all other users as interference. In general, the decoding order of the multicast and unicast messages at each receiver can be optimized according to the instantaneous channel condition. In this work, since the multicast message is intended for multiple users and should have a higher priority \cite{Kim_Superposition_Mag08,Simeone_LDM_GLOBECOM16}, we assume that the multicast message is decoded and subtracted before decoding the unicast message.
Thus, the signal-to-interference-plus-noise ratios (SINRs) of the multicast message and the unicast message at the $k$-th user are respectively expressed as
\begin{align}
\text{SINR}_{k}^{\text{M}} = \frac {\lvert \mathbf{h}_k^H \mathbf{w}_{0} \rvert^2} {\sum_{j = 1}^K \lvert \mathbf{h}_k^H \mathbf{w}_{j} \rvert^2 + \sigma_k^2}
\end{align}
and
\begin{align}
\text{SINR}_{k}^{\text{U}} = \frac {\lvert \mathbf{h}_k^H \mathbf{w}_{k} \rvert^2} {\sum_{j=1,\, j \neq k}^K \lvert \mathbf{h}_k^H \mathbf{w}_{j} \rvert^2 + \sigma_k^2}.
\end{align}

\subsection{Problem Formulation}
Our objective is to optimize the rate performance of both multicast and unicast services through joint design of the BS clustering scheme $\{s_{k,n}\}$ and the beamforming vectors $\{\mathbf{w}_k\}$ subject to the peak power and peak backhaul constraints on each individual BS. This is a multi-objective optimization problem. Thus we formulate a weighted sum of the unicast rate and the multicast rate maximization problem as follows: 
\begin{subequations} \label{pro:WSR-Clustering}
\begin{align} 
\mathcal{P}_{0}:~\mathop{\text{max}}_{\mathbf{w}, \mathbf{r}, \mathbf{s}} ~& \eta B r_{0} + (1 - \eta) B \sum_{k=1}^K r_{k} \label{obj:WSR-Clustering}\\ 
\text{s.t.} ~~
& \frac {\lvert \mathbf{h}_k^H \mathbf{w}_{0} \rvert^2} {\sum_{j = 1}^K \lvert \mathbf{h}_k^H \mathbf{w}_{j} \rvert^2 + \sigma_k^2} \geq 2^{r_{0}}-1, ~\forall~ k \in \mathcal{K}, \label{cons:WSR-Clustering-multicast-SINR}\\
& \frac {\lvert \mathbf{h}_k^H \mathbf{w}_{k} \rvert^2} {\sum_{j=1,\, j \neq k}^K \lvert \mathbf{h}_k^H \mathbf{w}_{j} \rvert^2 + \sigma_k^2} \geq 2^{r_{k}}-1, ~\forall~ k \in \mathcal{K}, \label{cons:WSR-Clustering-unicast-SINR}\\
& \sum_{k=0}^K \| \mathbf{w}_{k,n} \|_2^2 \leq P_n,~\forall~ n \in \mathcal{N}, \label{cons:WSR-Clustering-power} \\
& \lVert \mathbf{w}_{k,n} \rVert_2^2 \leq s_{k,n} P_n, ~\forall~ k \in \widehat{\mathcal{K}},~n \in \mathcal{N}, \label{cons:WSR-Clustering-cluster} \\
& \sum_{k=0}^K s_{k,n} B r_{k} \leq C_n,~\forall~ n \in \mathcal{N}, \label{cons:WSR-Clustering-backhaul} \\
& s_{k,n} \in \{0, 1\},~\forall~ k \in \widehat{\mathcal{K}},~n \in \mathcal{N}, \label{cons:WSR-Clustering-binary} 
\end{align} 
\end{subequations} 
where $r_{0}$ and $r_{k}$ are auxiliary variables which represent the transmission rates in bits/s/Hz of the multicast message and the $k$-th unicast message, respectively, $B$ is the available bandwidth of the wireless channel, and $\eta \in [0,1]$ is a weighting parameter between the multicast rate $R^{\text{M}} \triangleq B r_{0}$ and the unicast rate $R^{\text{U}} \triangleq B \sum_{k=1}^K r_{k}$.
For ease of notation, let $\mathbf{w} \triangleq \{\mathbf{w}_k \mid k \in \widehat{\mathcal{K}}\}$, $\mathbf{r} \triangleq \{r_k \mid k \in \widehat{\mathcal{K}}\}$, and $\mathbf{s} \triangleq \{s_{k,n} \mid k \in \widehat{\mathcal{K}}, n \in \mathcal{N}\}$. 

Note that besides the considered objective function, a more general form of weighted sum rate, e.g., $B \sum_{k=0}^K \omega_{k} r_{k}$, can be considered to account for possibly different priorities among all of the multicast and unicast services, where $\{\omega_{k}\}$ are weighting parameters that are determined by certain scheduling policy (e.g., proportional fair scheduler).

We also note that a minimum rate constraint for the multicast and each of the unicast services may be imposed to achieve certain quality of service, i.e., $r_{k} \geq r_k^{\text{min}}$ for all $k \in \widehat{\mathcal{K}}$ in practical systems. Such minimum rate constraints are all linear and hence do not change the structure of the problem (as well as the algorithm design). As such we do not consider the minimum rate constraints in problem $\mathcal{P}_{0}$ in order to fully characterize the multicast-unicast rate tradeoff.

By varying $\eta$, different priorities can be given to the multicast and the unicast services, and hence different achievable multicast-unicast rate pairs can be obtained. In the special case when $\eta = 0$, problem $\mathcal{P}_{0}$ reduces to 

\begin{subequations} \label{pro:TDM-unicast-WSR}
\begin{align} 
\mathcal{P}_{\text{U}}: \mathop{\text{max}}_{\{\mathbf{w}_k, r_{k}, s_{k,n}\}} ~& \sum_{k=1}^K r_{k} \\ 
\text{s.t.} \quad \quad
& \frac {\lvert \mathbf{h}_k^H \mathbf{w}_{k} \rvert^2} {\sum_{i=1, \, i \neq k}^K \lvert \mathbf{h}_k^H \mathbf{w}_{i} \rvert^2 + \sigma_k^2} \geq 2^{r_{k}}-1,~\forall~ k \in \mathcal{K}, \label{cons:TDM-unicast-WSR-SINR}\\
& \sum_{k=1}^K \| \mathbf{w}_{k,n} \|_2^2 \leq P_n,~\forall~ n \in \mathcal{N}, \label{cons:TDM-unicast-WSR-power} \\
& \lVert \mathbf{w}_{k,n} \rVert_2^2 \leq s_{k,n} P_n, ~\forall~ k \in \mathcal{K},~n \in \mathcal{N}, \label{cons:TDM-unicast-WSR-cluster} \\
& \sum_{k=1}^K s_{k,n} B r_{k} \leq C_n,~\forall~ n \in \mathcal{N}, \label{cons:TDM-unicast-WSR-backhaul} \\
& s_{k,n} \in \{0, 1\},~\forall~ k \in \mathcal{K},~n \in \mathcal{N}, \label{cons:TDM-unicast-WSR-binary} 
\end{align} 
\end{subequations}
which is equivalent to the sparse unicast beamforming design problem in \cite{WeiYu_access}, where the binary BS clustering variable $s_{k,n}$ is replaced by the indicator function $\mathbbm{1} \left \{\lVert \mathbf{w}_{k,n} \rVert_2^2 \right \}$.
When $\eta = 1$, problem $\mathcal{P}_{0}$ reduces to a pure multicast beamforming design problem:
\begin{subequations} \label{pro:TDM-multicast-WSR}
\begin{align} 
\mathcal{P}_{\text{M}}: ~\mathop{\text{max}}_{\{\mathbf{w}_{0}, r_{0}, s_{0,n}\}} ~& r_{0} \\ 
\text{s.t.} \quad \quad
& \frac {\lvert \mathbf{h}_k^H \mathbf{w}_{0} \rvert^2} {\sigma_k^2} \geq 2^{r_{0}}-1,~\forall~ k \in \mathcal{K}, \label{cons:TDM-multicast-WSR-SINR}\\
& \lVert \mathbf{w}_{0,n} \rVert_2^2 \leq s_{0,n} P_n, ~\forall~ n \in \mathcal{N}, \label{cons:TDM-multicast-WSR-cluster} \\
& r_{0,n} B r_{0} \leq C_n,~\forall~ n \in \mathcal{N}, \label{cons:TDM-multicast-WSR-backhaul} \\
& s_{0,n} \in \{0, 1\},~\forall~ n \in \mathcal{N}. \label{cons:TDM-multicast-WSR-binary} 
\end{align} 
\end{subequations}

Problem $\mathcal{P}_{0}$ is a non-convex MINLP problem \cite{Burer_nonconvex-MINLP_12}, which is NP-hard in general. 
Obtaining its optimal solution is challenging due to the non-convexity of the SINR constraints \eqref{cons:WSR-Clustering-multicast-SINR} and \eqref{cons:WSR-Clustering-unicast-SINR}, the combinatorial nature of the BS clustering variable $s_{k,n}$ in \eqref{cons:WSR-Clustering-binary}, and the coupling between the variables $s_{k,n}$ and $r_{k}$ in the backhaul constraint \eqref{cons:WSR-Clustering-backhaul}. Even when the BS clustering scheme $\mathbf{s}$ is given, $\mathcal{P}_{0}$ is still non-convex and computationally difficult. 
In the following sections, we first develop a BB-based algorithm to find the global optimum of problem $\mathcal{P}_{0}$. We then propose a low-complexity algorithm to find a high-quality approximate solution. Both of the proposed algorithms can also be applied to problems $\mathcal{P}_{\text{U}}$ and $\mathcal{P}_{\text{M}}$.

\section{BB-based Optimal Algorithm} \label{sec:BB-based Optimal Algorithm}
In this section, we propose a global optimal algorithm to solve problem $\mathcal{P}_{0}$ based on the BB method. 

\subsection{Overview of the BB Method}
The BB method is a general framework for designing global optimization algorithms for non-convex problems. The BB method is non-heuristic in the sense that it generates a sequence of asymptotically tight upper and lower bounds on the optimal objective value; it terminates with a certificate proving that the found point is $\epsilon$-optimal \cite{Boyd_BB_note_11}. 

A BB algorithm consists of a systematic enumeration procedure, which recursively partitions the feasible region of the original problem into smaller subregions and constructs subproblems over the partitioned subregions. 
An upper (for solving a maximization problem) bound for each subproblem is often computed by solving a convex relaxation problem defined over the corresponding subregion; a lower bound is obtained from the best known feasible solution generated by the enumeration procedure or by some other heuristic or local algorithms. 
A subproblem is discarded if it cannot produce a better solution than the best one found so far by the algorithm.
The performance of the BB algorithm depends on the efficient estimation of the lower and upper bounds of each subproblem. To ensure the convergence, the bounds should become tight as the number of subregions in the partition grows. 

Recently, the BB method has been used for beamforming design in cellular networks. For example, a customized BB algorithm is proposed in \cite{Liu_optimalmulticast_TSP17} for single-group multicast beamforming and then extended in \cite{Liu_SPAWC17} for joint multicast and unicast beamforming. A monotonic optimization based branch-and-reduce-and-bound (BRB) algorithm is proposed in \cite{Tervo_optimal-beamforming_TSP15} to solve the energy efficiency maximization problem in a multiuser MISO downlink system. The BRB algorithm is then extended in \cite{Luong_optimal-beamforming_TSP17} for joint remote radio head selection and beamforming design in cloud radio access networks.



\subsection{Convex Relaxations}
In this subsection, we introduce some effective convex relaxations for the non-convex constraints of $\mathcal{P}_{0}$, which play an important role in finding the lower and upper bounds in the proposed BB-based algorithm for solving the problem. 

Define $g_k(\mathbf{w}) \triangleq \sum_{i = 1}^K \lvert \mathbf{h}_k^H \mathbf{w}_{i} \rvert^2 + \sigma_k^2$, then without loss of optimality the unicast SINR constraint \eqref{cons:WSR-Clustering-unicast-SINR} can be rewritten as 
\begin{align} \label{cons:WSR-Clustering-unicast-SINR-SOC}
\mathbf{h}_k^H \mathbf{w}_{k} \geq \sqrt{(2^{r_{k}}-1)/2^{r_{k}}} \sqrt{g_k(\mathbf{w})},~
\Im \{\mathbf{h}_k^H \mathbf{w}_{k}\} = 0, ~\forall~ k \in \mathcal{K},
\end{align}
which is a convex second-order cone (SOC) constraint when $\{r_{k}\}$ is given. 
For the multicast SINR constraint \eqref{cons:WSR-Clustering-multicast-SINR}, since all the users share the same multicast beamformer $\mathbf{w}_{0}$ and the channel vectors $\{\mathbf{h}_k\}$ are linearly independent, there is only one user's multicast SINR constraint (assume without loss of generality it is the $K$-th user) can be rewritten into the convex SOC form when $r_{0}$ is given, i.e.,
\begin{align} \label{cons:WSR-Clustering-multicast-SINR-SOC-K}
\mathbf{h}_K^H \mathbf{w}_{0} \geq \sqrt{2^{r_{0}}-1} \sqrt{g_K(\mathbf{w})},~\Im \{\mathbf{h}_K^H \mathbf{w}_{0}\} = 0.
\end{align}
The rest can be represented as
\begin{align} \label{cons:WSR-Clustering-multicast-SINR-SOC}
\lvert \mathbf{h}_k^H \mathbf{w}_{0} \rvert \geq \sqrt{2^{r_{0}}-1} \sqrt{g_k(\mathbf{w})}, ~\forall~ k \in \mathcal{K} \setminus \{K\},
\end{align}
which is non-convex. The above transformations \eqref{cons:WSR-Clustering-multicast-SINR-SOC-K} and \eqref{cons:WSR-Clustering-multicast-SINR-SOC} have also been used in \cite{Liu_SPAWC17} to facilitate the joint design of multicast and unicast beamforming.

Next, we present convex relaxations for the non-convex constraints \eqref{cons:WSR-Clustering-multicast-SINR-SOC} and \eqref{cons:WSR-Clustering-backhaul} in the following propositions. 

\begin{proposition}[\cite{Liu_SPAWC17}, Proposition 1] \label{prop:Convex-envelope-multicast}
Let $\varphi_{k}$ be the argument of $\mathbf{h}_k^H \mathbf{w}_{0}$, where $\varphi_{k} \in [\underline{\varphi}_{k}, \bar{\varphi}_{k}]$, $0 \leq \underline{\varphi}_{k} \leq \bar{\varphi}_{k} < 2\pi$, and $\mathcal{D}_{k}^{[\underline{\varphi}_{k}, \bar{\varphi}_{k}]}(r_{0})$ denote the set of $\mathbf{w}$ defined by the inequality $\lvert \mathbf{h}_k^H \mathbf{w}_{0} \rvert \geq \sqrt{2^{r_{0}}-1} \sqrt{g_k(\mathbf{w})}$ for a given $r_{0}$, for all $k \in \mathcal{K} \setminus \{K\}$. Suppose that $\bar{\varphi}_{k} - \underline{\varphi}_{k} \leq \pi$, then the convex envelope of $\mathcal{D}_{k}^{[\underline{\varphi}_{k}, \bar{\varphi}_{k}]}(r_{0})$ is given by
\begin{align} \label{cons:WSR-Clustering-multicast-SINR-SOC-envelope}
Conv(\mathcal{D}_{k}^{[\underline{\varphi}_{k}, \bar{\varphi}_{k}]}(r_{0})) =& \left \{ \mathbf{w} \mid \sin(\underline{\varphi}_{k}) \Re\{\mathbf{h}_k^H \mathbf{w}_{0}\} - \cos(\underline{\varphi}_{k}) \Im\{\mathbf{h}_k^H \mathbf{w}_{0}\} \leq 0, \right. \nonumber\\
&\qquad \left. \sin(\bar{\varphi}_{k}) \Re\{\mathbf{h}_k^H \mathbf{w}_{0}\} - \cos(\bar{\varphi}_{k}) \Im\{\mathbf{h}_k^H \mathbf{w}_{0}\} \geq 0, \right. \nonumber \\
&\qquad \left. x_{k} \Re\{\mathbf{h}_k^H \mathbf{w}_{0}\} + y_{k} \Im\{\mathbf{h}_k^H \mathbf{w}_{0}\}  \geq (x_{k}^2 + y_{k}^2) \sqrt{2^{r_{0}}-1} \sqrt{g_k(\mathbf{w})} \right\},
\end{align} 

where $x_{k} = (\cos(\underline{\varphi}_{k}) + \cos(\bar{\varphi}_{k})) / 2$ and $y_{k} = (\sin(\underline{\varphi}_{k}) + \sin(\bar{\varphi}_{k})) / 2$. 
\end{proposition}

It is easy to verify that the smaller the width of the interval $[\underline{\varphi}_{k}, \bar{\varphi}_{k}]$, the tighter the convex envelope. As $\bar{\varphi}_{k} - \underline{\varphi}_{k}$ goes to zero, the set $Conv(\mathcal{D}_{k}^{[\underline{\varphi}_{k}, \bar{\varphi}_{k}]}(r_{0}))$ becomes $\mathcal{D}_{k}^{[\underline{\varphi}_{k}, \bar{\varphi}_{k}]}(r_{0})$ and the convex envelope becomes tight.

\begin{proposition}[\cite{Biconvex83}, Theorem 2] \label{prop:Convex-envelope-bakchaul} 
Suppose that $\Omega = \{(\mathbf{x}, \mathbf{y}) \mid \mathbf{x}, \mathbf{y} \in \mathbb{R}^n, \underline{\mathbf{x}}\leq \mathbf{x} \leq \bar{\mathbf{x}}, \underline{\mathbf{y}} \leq \mathbf{y} \leq \bar{\mathbf{y}}\}$, then the convex envelope of function $f = \mathbf{x}^T\mathbf{y}$ over $\Omega$ is given by
\begin{equation} \label{cons:WSR-backhaul-bilinear-envelope}
Conv_{\Omega}(\mathbf{x}^T\mathbf{y}) = \sum_{i=1}^n \max \{ \underline{y}_i x_i + \underline{x}_i y_i - \underline{x}_i \underline{y}_i, \bar{y}_i x_i + \bar{x}_i y_i - \bar{x}_i \bar{y}_i \}.
\end{equation}  
\end{proposition}

Recall that the convex envelope of a function $f$ over a set $\Omega$ is the pointwise supremum of all convex functions which underestimate $f$ over $\Omega$ \cite{Biconvex83}, i.e., $Conv_{\Omega}(\mathbf{x}^T\mathbf{y})$ is convex and $Conv_{\Omega}(\mathbf{x}^T\mathbf{y}) \leq \mathbf{x}^T\mathbf{y}$ over $\Omega$. It is easy to see that when the box region $\Omega$ shrinks to a point, the convex envelope $Conv_{\Omega}(\mathbf{x}^T\mathbf{y})$ becomes tight.

\subsection{Proposed BB-based Algorithm} \label{sec:Proposed BB-based Algorithm}
For ease of the presentation, let $\mathbf{q} = [\mathbf{s}^T, \mathbf{r}^T, \bm{\varphi}^T]^T \in \mathbb{R}_{+}^{N_q}$ be the variable vector of interest where $N_q = (K+1)N + (K+1) + (K-1)$. Here the binary variable $\mathbf{s}$ is relaxed to be continuous. Notice that $\mathbf{q}$ belongs to the box $\mathcal{Q}_{\text{init}} = [\underline{\mathbf{q}}, \bar{\mathbf{q}}]$, where the lower and upper vertices are given by
\begin{align} 
\underline{\mathbf{q}} = \mathbf{0}_{N_q},~\bar{\mathbf{q}} = [\mathbf{1}_{(K+1)N}^T, \mathbf{r}_{\text{max}}^T, 2\pi \times \mathbf{1}_{(K-1)}^T]^T. \nonumber
\end{align}
Here, $\mathbf{r}_{\text{max}}$ is an upper bound of the rate $\mathbf{r}$, each element of which can be obtained by transmitting the total available power $P_{\text{total}} \triangleq \sum_{n = 1}^N P_n$ towards a single user and cannot exceed the maximum backhaul capacity of the BSs. In specific, we have $r_{\text{max}}^k = \min \{\max_{n \in \mathcal{N}} \{{C_n}\}, \log_2(1 + P_{\text{total}} \|\mathbf{h}_k \|_2^2 / \sigma_k^2)\}$, for all $k \in \mathcal{K}$, and $r_{\text{max}}^0 = \min_{k \in \mathcal{K}} \{r_{\text{max}}^k\}$. 

Let $\mathcal{R}^t$, $\Phi_{\text{U}}^t$, and $\Phi_{\text{L}}^t$ denote the box list, the upper bound, and the lower bound of the optimal objective value of the original problem $\mathcal{P}_{0}$ at the $t$-th iteration, respectively. Let $\Phi_{\text{U}}(\mathcal{Q})$ and $\Phi_{\text{L}}(\mathcal{Q})$ denote the upper bound and the lower bound of the objective value over a given box region $\mathcal{Q}$. The proposed BB algorithm works as follows:

\subsubsection{Branch} 
At the $t$-th iteration, we select a box in $\mathcal{R}^t$ and split it into two smaller ones. An effective method for selecting the candidate box is to choose the one with the largest upper bound, i.e., $\mathcal{Q}^{*} = \arg\max_{\mathcal{Q} \in \mathcal{R}^t} \Phi_{\text{U}}(\mathcal{Q})$. The selected box $\mathcal{Q}^{*} = [\mathbf{a}, \mathbf{b}]$ is then split along the longest edge, e.g., $j^{*} = \arg\max_{1\leq j\leq N_q} \{b_j - a_j\}$, to create two boxes with equal size
\begin{equation} \label{cons:WSR-Clustering-splitting-rule}
\begin{array}{l} 
\mathcal{Q}_{(1)}^{*} = 
\begin{cases}
[\mathbf{a}, \mathbf{b} - \mathbf{e}_{j^{*}}], &\text{if } j^{*} \leq (K+1)N, \\
[\mathbf{a}, \mathbf{b} - (b_{j^{*}} - a_{j^{*}})/2 \times \mathbf{e}_{j^{*}}], &\text{otherwise}, 
\end{cases}  \\
\mathcal{Q}_{(2)}^{*} = 
\begin{cases}
[\mathbf{a} + \mathbf{e}_{j^{*}}, \mathbf{b}], &\text{if } j^{*} \leq (K+1)N, \\
[\mathbf{a} + (b_{j^{*}} - a_{j^{*}})/2 \times \mathbf{e}_{j^{*}}, \mathbf{b}], &\text{otherwise},
\end{cases} 
\end{array}
\end{equation}
where $\mathbf{e}_{j^{*}}$ is the $j^{*}$-th standard basis vector. Note that the above splitting rule takes the binary variable $\mathbf{s}$ into account, which is adjusted to be in the Boolean set. 

\subsubsection{Bound}  
The bounding operation is to compute the upper and lower bounds over the newly added box $\mathcal{Q}$, $\mathcal{Q} \in \{\mathcal{Q}_{(1)}^{*}, \mathcal{Q}_{(2)}^{*} \}$, and update the upper bound $\Phi_{\text{U}}^{t+1}$ and the lower bound $\Phi_{\text{L}}^{t+1}$.

{\em Upper Bound:} The upper bound $\Phi_{\text{U}}(\mathcal{Q})$ is computed by solving a convex relaxation of problem $\mathcal{P}_{0}$ over the box $\mathcal{Q}$. 

The SINR constraints \eqref{cons:WSR-Clustering-multicast-SINR}, \eqref{cons:WSR-Clustering-unicast-SINR} can be transformed into constraints \eqref{cons:WSR-Clustering-unicast-SINR-SOC}, \eqref{cons:WSR-Clustering-multicast-SINR-SOC-K}, and \eqref{cons:WSR-Clustering-multicast-SINR-SOC}, which are still non-convex. We first deal with constraints \eqref{cons:WSR-Clustering-unicast-SINR-SOC} and \eqref{cons:WSR-Clustering-multicast-SINR-SOC-K} by relaxing them as 
\begin{align} \label{cons:WSR-Clustering-unicast-SINR-SOC-relaxed}
\mathbf{h}_k^H \mathbf{w}_{k} \geq \sqrt{(2^{\underline{r}_{k}}-1)/(2^{\underline{r}_{k}})} \sqrt{g_k(\mathbf{w})},~
\Im \{\mathbf{h}_k^H \mathbf{w}_{k}\} = 0, ~\forall~ k \in \mathcal{K} 
\end{align}
and
\begin{align} \label{cons:WSR-Clustering-multicast-SINR-SOC-K-relaxed}
\mathbf{h}_K^H \mathbf{w}_{0} \geq \sqrt{(2^{\underline{r}_{0}}-1)} \sqrt{g_K(\mathbf{w})},~ \Im \{\mathbf{h}_K^H \mathbf{w}_{0}\} = 0,
\end{align}
respectively. Then we replace constraint \eqref{cons:WSR-Clustering-multicast-SINR-SOC} by its convex envelope with the given $\underline{r}_{0}$ according to Proposition \ref{prop:Convex-envelope-multicast}
\begin{align} \label{cons:WSR-Clustering-multicast-SINR-SOC-relaxed}
\mathbf{w} \in Conv(\mathcal{D}_{k}^{[\underline{\varphi}_{k}, \bar{\varphi}_{k}]}(\underline{r}_{0})), ~\forall~ k \in \mathcal{K} \setminus \{K\}.
\end{align}
Note that the convex envelope only takes effect when $\bar{\varphi}_{k} - \underline{\varphi}_{k} \leq \pi$. If there is any user $k$ such that $\bar{\varphi}_{k} - \underline{\varphi}_{k} > \pi$, it means $\mathbf{h}_k^H \mathbf{w}_{0}$ can take value of the whole complex plane, and we just remove the multicast SINR constraint of user $k$ from \eqref{cons:WSR-Clustering-multicast-SINR-SOC-relaxed}.

For the non-convex backhaul constraint \eqref{cons:WSR-Clustering-backhaul}, we can relax it into
\begin{align} \label{cons:WSR-Clustering-backhaul-relaxed}
\sum_{k=0}^K B \max \{ \underline{r}_k s_{k,n} + \underline{s}_{k,n} r_k - \underline{s}_{k,n} \underline{r}_k, \bar{r}_k s_{k,n} + \bar{s}_{k,n} r_k - \bar{s}_{k,n} \bar{r}_k \} \leq C_n,~\forall~ n \in \mathcal{N}, 
\end{align}
according to Proposition \ref{prop:Convex-envelope-bakchaul}.

In addition, since $\mathbf{q}$ is restricted within the box $\mathcal{Q}$, we have
\begin{align} \label{cons:WSR-Clustering-binary-relaxed}
\underline{r}_{k} \leq r_{k} \leq \bar{r}_{k}, ~\underline{s}_{k,n} \leq s_{k,n} \leq \bar{s}_{k,n},~\forall~ k \in \widehat{\mathcal{K}}, n \in \mathcal{N}.
\end{align}

Note that the current form of constraint \eqref{cons:WSR-Clustering-cluster} may produce a loose relaxation when the binary variable $s_{k,n}$ is relaxed to be a continuous one, since $\lVert \mathbf{w}_{k,n} \rVert_2^2$ can be possibly much smaller than $P_n$. To tight the relaxation, we adopt the perspective reformulation in \cite{perspective2010,Tervo_optimal-beamforming_TSP15} to rewrite constraints \eqref{cons:WSR-Clustering-power} and \eqref{cons:WSR-Clustering-cluster} into the following form:
\begin{align} 
& \sum_{k=0}^K v_{k,n} \leq P_n,~\forall~ n \in \mathcal{N}, \label{cons:WSR-Clustering-power-relaxed} \\
& \lVert \mathbf{w}_{k,n} \rVert_2^2 \leq s_{k,n} v_{k,n}, ~\forall~ k \in \widehat{\mathcal{K}},~n \in \mathcal{N}, \label{cons:WSR-Clustering-cluster-relaxed} 
\end{align}
where $v_{k,n}$ can be interpreted as a soft power level for the $n$-th BS serving the $k$-th message and is optimized under the power constraint \eqref{cons:WSR-Clustering-power-relaxed}. Further, constraint \eqref{cons:WSR-Clustering-cluster-relaxed} can be rewritten as
\begin{align} 
\lVert \mathbf{w}_{k,n}^T, \frac{1}{2}(s_{k,n} - v_{k,n}) \rVert_2 \leq \frac{1}{2}(s_{k,n} + v_{k,n}), ~\forall~ k \in \widehat{\mathcal{K}},~n \in \mathcal{N}, \label{cons:WSR-Clustering-cluster-relaxed-SOC} 
\end{align}
which is an SOC constraint when $s_{k,n}$ is relaxed to be continuous. 

Finally, we can obtain $\Phi_{\text{U}}(\mathcal{Q})$ by solving the following relaxed problem:
\begin{align} \label{pro:WSR-Clustering-Upper}
\mathop{\text{max}}_{\mathbf{w}, \mathbf{r}, \mathbf{s}} ~& \eta B r_{0} + (1 - \eta) B \sum_{k=1}^K r_{k} \\ 
\text{s.t.} ~~ &\eqref{cons:WSR-Clustering-unicast-SINR-SOC-relaxed},\eqref{cons:WSR-Clustering-multicast-SINR-SOC-K-relaxed},\eqref{cons:WSR-Clustering-multicast-SINR-SOC-relaxed},\eqref{cons:WSR-Clustering-backhaul-relaxed},\eqref{cons:WSR-Clustering-binary-relaxed},\eqref{cons:WSR-Clustering-power-relaxed}, \text{and}~\eqref{cons:WSR-Clustering-cluster-relaxed-SOC}. \nonumber
\end{align} 
Problem \eqref{pro:WSR-Clustering-Upper} is a convex problem, which can be equivalently reformulated as a second-order cone programming (SOCP) and efficiently solved using a general-purpose solver via interior-point methods \cite{Boyd_convex_optimization}. Note that problem \eqref{pro:WSR-Clustering-Upper} may be infeasible. If this happens, it indicates that the box $\mathcal{Q}$ does not contain the optimal solution and we just set $\Phi_{\text{U}}(\mathcal{Q})$ and $\Phi_{\text{L}}(\mathcal{Q})$ as $-\infty$.

After obtaining the upper bounds $\Phi_{\text{U}}(\mathcal{Q})$, for $\mathcal{Q} \in \{\mathcal{Q}_{(1)}^{*}, \mathcal{Q}_{(2)}^{*} \}$, we can form $\mathcal{R}^{t+1}$ by removing $\mathcal{Q}^{*}$ from $\mathcal{R}^{t}$ and adding $\mathcal{Q}^{*}_{(1)}$ and $\mathcal{Q}^{*}_{(2)}$ if their upper bounds are larger than or equal to the current best lower bound $\Phi_{\text{L}}^{t}$, i.e., $\mathcal{R}^{t+1} = \mathcal{R}^{t} \setminus \{\mathcal{Q}^{*}\} \cup \{\mathcal{Q}^{*}_{(i)} \mid \Phi_{\text{U}}(\mathcal{Q}^{*}_{(i)}) \geq \Phi_{\text{L}}^{t}, i = 1, 2\}$. Note that the maximum of the upper bounds over all boxes in $\mathcal{R}^{t+1}$ is an upper bound of the optimal objective value of the original problem. Therefore, we update the upper bound as $\Phi_{\text{U}}^{t+1} = \max_{\mathcal{Q} \in \mathcal{R}^{t+1}} \Phi_{\text{U}}(\mathcal{Q})$.

{\em Lower Bound:} To obtain a lower bound, we need to find a feasible solution of the original problem $\mathcal{P}_{0}$. This can be done by gaining some insights from the optimal solution of problem \eqref{pro:WSR-Clustering-Upper}. 

After obtaining beamforming vector $\{\mathbf{w}_{k,n}^*\}$ of problem \eqref{pro:WSR-Clustering-Upper}, we can turn off some data links with small transmit power and keep the other ones active, i.e., force $s_{k,n} = 0$ if $\|\mathbf{w}_{k,n}^* \|_2^2$ is small enough and set the remaining $s_{k,n} = 1$. Since the data link with a lower power gain contributes less to the weighted sum rate and should have a higher priority to be turned off. Denote $p_j$ as the $j$-th largest element of $\{\|\mathbf{w}_{k,n}^*\|_2^2\}$. Let
\begin{align}  \label{equ:BB-lower-bound-solution}
\tilde{\mathbf{w}}_{k,n} = 
	\begin{cases}
	\mathbf{0}_L, &\text{if } \|\mathbf{w}_{k,n}^* \|_2^2 < p_j, \\
	\mathbf{w}_{k,n}^*, &\text{otherwise}, 
	\end{cases}  
~\text{and}~~
\tilde{s}_{k,n} = 
	\begin{cases}
	0, &\text{if } \|\mathbf{w}_{k,n}^*\|_2^2 < p_j, \\
	1, &\text{otherwise}.
	\end{cases} 
\end{align} 
Then, we can calculate the multicast rate $r_{0}(\tilde{\mathbf{w}})$ and unicast rate $r_{k}(\tilde{\mathbf{w}})$ as
\begin{align} \label{equ:BB-lower-bound-solution-R0}
r_{0}(\tilde{\mathbf{w}}) = \min_{k \in \mathcal{K}} \log_2 \left(1 + \frac {\lvert \mathbf{h}_k^H \tilde{\mathbf{w}}_{0} \rvert^2} {\sum_{i = 1}^K \lvert \mathbf{h}_k^H \tilde{\mathbf{w}}_{i} \rvert^2 + \sigma_k^2} \right)
\end{align}
and 
\begin{align} \label{equ:BB-lower-bound-solution-Rk}
r_{k}(\tilde{\mathbf{w}}) = \log_2 \left(1 + \frac {\lvert \mathbf{h}_k^H \tilde{\mathbf{w}}_{k} \rvert^2} {\sum_{i=1, \, i \neq k}^K \lvert \mathbf{h}_k^H \tilde{\mathbf{w}}_{i} \rvert^2 + \sigma_k^2}\right),~\forall~ k \in \mathcal{K},
\end{align}
respectively.

If the backhaul constraint \eqref{cons:WSR-Clustering-backhaul} is satisfied, i.e., $\sum_{k=0}^K \tilde{s}_{k,n} r_{k}(\tilde{\mathbf{w}}) \leq C_n$, for all $n \in \mathcal{N}$, then $\{\tilde{\mathbf{w}}, \tilde{\mathbf{s}}, r_{k}(\tilde{\mathbf{w}}) \}$ itself is a feasible solution of the original problem $\mathcal{P}_{0}$. Otherwise, we can scale $\{r_{k}(\tilde{\mathbf{w}})\}$ to be feasible. Therefore, a feasible solution of problem $\mathcal{P}_{0}$ is given by $\{\tilde{\mathbf{w}}, \tilde{s}, \tilde{r}_{k}\}$ where
\begin{align} \label{equ:BB-lower-bound-solution-r}
\tilde{r}_{k} = \min  \left\{ \min_{n \in \mathcal{N}} \left \{\frac {C_n} {\sum_{k=0}^K \tilde{s}_{k,n} r_{k}(\tilde{\mathbf{w}})} \right \}, 1 \right \} r_{k}(\tilde{\mathbf{w}}),~\forall~ k \in \widehat{\mathcal{K}}.
\end{align}
Note that for each $j \in \{1,2,\dots, (K+1)N\}$, we can find such a feasible solution $\{\tilde{\mathbf{w}}, \tilde{s}, \tilde{r}_{k}\}$ and its corresponding objective $\Phi^{j}_{\text{L}}(\mathcal{Q}) = \eta K B \tilde{r}_{0} + (1 - \eta) \sum_{k=1}^K B \tilde{r}_{k}$. The lower bound $\Phi_{\text{L}}(\mathcal{Q})$ can be obtained by finding the best $j$, which yields the largest objective among all feasible solutions, i.e.,
\begin{align} \label{equ:BB-lower-bound}
\Phi_{\text{L}}(\mathcal{Q}) = \max_{j \in \{1,2,\dots, (K+1)N\}} \{\Phi^{j}_{\text{L}}(\mathcal{Q})\}.
\end{align}

Finally, we can obtain a better lower bound of the optimal objective value of the original problem if the lower bounds of the newly added boxes can provide a larger lower bound than that of the previous iteration, i.e., $\Phi_{\text{L}}^{t+1} = \max \{\Phi_{\text{L}}(\mathcal{Q}^{*}_{(1)}), \Phi_{\text{L}}(\mathcal{Q}^{*}_{(2)}), \Phi_{\text{L}}^{t}\}$.

The overall BB-based algorithm for solving problem $\mathcal{P}_{0}$ is summarized in Alg. \ref{alg:WSR-WSR-Clustering-BB}.
\begin{algorithm}[htt] 
\caption{The BB-based algorithm for globally solving problem $\mathcal{P}_{0}$} \label{alg:WSR-WSR-Clustering-BB}
\begin{algorithmic}[0]
\STATE \textbf{Initialization:} Initialize $\mathcal{R}^0 \leftarrow \{\mathcal{Q}_{\text{init}}\}$ and the iteration index $t \leftarrow 0$. Find the upper bound $\Phi_{\text{U}}(\mathcal{Q}_{\text{init}})$ by solving problem \eqref{pro:WSR-Clustering-Upper}, and the lower bound $\Phi_{\text{L}}(\mathcal{Q}_{\text{init}})$ according to \eqref{equ:BB-lower-bound}. Set $\Phi_{\text{L}}^0 = \Phi_{\text{L}}(\mathcal{Q}_{\text{init}})$, $\Phi_{\text{U}}^0 = \Phi_{\text{U}}(\mathcal{Q}_{\text{init}})$, and the tolerance $\epsilon$.
\STATE \textbf{While} $\Phi_{\text{U}}^{t} - \Phi_{\text{L}}^{t} > \epsilon $
\begin{enumerate}
	\item {\bf Branch:} Select the box $\mathcal{Q}^{*}$ in $\mathcal{R}^{t}$ with the largest upper bound, i.e., $\Phi_{\text{U}}(\mathcal{Q}^{*}) = \Phi_{\text{U}}^t$, and split it into two boxes $\mathcal{Q}^{*}_{(1)}$ and $\mathcal{Q}^{*}_{(2)}$ according to the splitting rule \eqref{cons:WSR-Clustering-splitting-rule}.
	\item {\bf Bound:} For each box $\mathcal{Q}^{*}_{(i)}~(i=1,2)$, find its upper bound $\Phi_{\text{U}}(\mathcal{Q}^{*}_{(i)})$ by solving problem \eqref{pro:WSR-Clustering-Upper} and its lower bound $\Phi_{\text{L}}(\mathcal{Q}^{*}_{(i)})$ according to \eqref{equ:BB-lower-bound}.
	\item Update $\mathcal{R}^{t+1} = \mathcal{R}^{t} \setminus \{\mathcal{Q}^{*}\} \cup \{\mathcal{Q}^{*}_{(i)} \mid \Phi_{\text{U}}(\mathcal{Q}^{*}_{(i)}) \geq \Phi_{\text{L}}^{t}, i = 1, 2\}$.
	\item Update $\Phi_{\text{U}}^{t+1} = \max_{\mathcal{Q} \in \mathcal{R}^{t+1}} \Phi_{\text{U}}(\mathcal{Q})$.
	\item Update $\Phi_{\text{L}}^{t+1} = \max \{\Phi_{\text{L}}(\mathcal{Q}^{*}_{(1)}), \Phi_{\text{L}}(\mathcal{Q}^{*}_{(2)}), \Phi_{\text{L}}^{t}\}$.
	\item Set $t \leftarrow t + 1$.	
\end{enumerate}
\STATE \textbf{End}
\end{algorithmic}
\end{algorithm}

\subsection{Convergence and Complexity Analysis}
\subsubsection{Convergence}
One important condition for the convergence of the BB-based algorithm is that the upper and lower bounds over a box region become tight as the box shrinks to a point. More precisely, as the length of the longest edge of the box $\mathcal{Q}$, denoted by $\text{size}(\mathcal{Q})$, goes to zero, the gap between upper and lower bounds converges to zero. We formally summarize the result in the following lemma:
\begin{lemma} \label{lemma_convergence_BB}
For any given $\epsilon > 0$ and $\mathcal{Q} \subseteq \mathcal{Q}_{\text{init}}$, there exists a $\delta \in (0,1)$ such that $KB\delta -2\eta KB\log_2 \left(\cos(\delta/2) \right) \leq \epsilon$ and when $\text{size}(\mathcal{Q}) \leq \delta$, we have $\Phi_{\text{U}}(\mathcal{Q}) - \Phi_{\text{L}}(\mathcal{Q}) \leq \epsilon$.
\end{lemma}
\begin{IEEEproof}
See Appendix \ref{proof_of_lemma_convergence_BB}.
\end{IEEEproof}

Lemma \ref{lemma_convergence_BB} indicates that for any given tolerance $\epsilon$, we can always find an $\epsilon$-optimal solution when the size of the box is sufficiently small. Note that by adopting the splitting rule \eqref{cons:WSR-Clustering-splitting-rule}, the size of the selected $\mathcal{Q}^{*}$ at the iteration of Alg. \ref{alg:WSR-WSR-Clustering-BB} converges to zero, i.e., $\text{size}(\mathcal{Q}^{*}) \to 0$. The proof is provided in \cite{Boyd_BB_note_11} and we omit it here for brevity.

\subsubsection{Complexity Analysis}
In Alg. \ref{alg:WSR-WSR-Clustering-BB}, the most computationally expensive part is to calculate the upper and lower bounds in Step 2). Obtaining the upper bound requires solving an SOCP problem in the form of \eqref{pro:WSR-Clustering-Upper}, and its worst-case computational complexity is approximately $\mathcal{O}((KNL)^{3.5})$ by adopting the interior-point methods \cite{Boyd_SOCP_1998}. Obtaining the lower bound in \eqref{equ:BB-lower-bound} takes at most $(K+1)N$ times; each has a complexity of $\mathcal{O}(K^2NL)$, which mainly lies in the rate computation in \eqref{equ:BB-lower-bound-solution-R0} and \eqref{equ:BB-lower-bound-solution-Rk}. Therefore, the computational complexity of Alg. \ref{alg:WSR-WSR-Clustering-BB} at each iteration mainly comes from calculating the upper bound in Step 2). Regarding to the maximum iteration number of Alg. \ref{alg:WSR-WSR-Clustering-BB}, we have the following lemma:
\begin{lemma} \label{lemma_complexity_BB}
For any given small constant $\epsilon > 0$ and any instance of problem $\mathcal{P}_{0}$, the proposed BB-based algorithm will return an $\epsilon$-optimal solution within at most
\begin{align} \label{equ:maximum_iteration}
T_{\text{max}}^B := 2^{(K+1)N} \left \lceil \left(\frac{2\pi}{\delta/2} \right)^{K-1} \prod_{k=0}^{K} \frac{r_{\text{max}}^k}{\delta/2}  \right \rceil
\end{align}
iterations, where $\delta = g^{-1}(\epsilon)$ is the inverse function of $g(\delta) = KB\delta -2\eta KB\log_2 \left(\cos(\delta/2) \right)$.
\end{lemma}
\begin{IEEEproof}
See Appendix \ref{proof_of_lemma_complexity_BB}.
\end{IEEEproof}

Since Alg. \ref{alg:WSR-WSR-Clustering-BB} requires at most $T_{\text{max}}^B$ iterations to converge, the worst-case computational complexity of Alg. \ref{alg:WSR-WSR-Clustering-BB} is therefore $\mathcal{O}(T_{\text{max}}^B(KNL)^{3.5})$.
As we can see from Lemma \ref{lemma_complexity_BB}, $T_{\text{max}}^B$ can be very large if the tolerance $\epsilon$ is small. Nevertheless, the proposed BB-based algorithm can be used as the network performance benchmark. Considering the practical implementation, in the next section, we will propose a low-complexity algorithm through sparse beamforming design. Simulation results show that it can achieve high performance that is very close to the optimum.

\section{CCP-based Low-Complexity Algorithm} \label{sec:CCP-based Low-complexity Algorithm}
In this section, we reformulate the joint BS clustering and beamforming problem as an equivalent sparse beamforming design problem and propose a CCP-based low-complexity algorithm to solve it approximately. 

\subsection{Sparse Beamforming Reformulation}
Recall that $\| \mathbf{w}_{k,n} \|_2^2 = 0$ if $s_{k,n} = 0$. Without loss of optimality, the binary BS clustering variable $s_{k,n}$ can be replaced by $\left \lVert \| \mathbf{w}_{k,n} \|_2^2 \right \rVert_0$, as in \cite{content_centric_TWC16,WeiYu_access}. Therefore, $\mathcal{P}_{0}$ can be rewritten as
\begin{subequations} \label{pro:WSR}
\begin{align} 
\mathcal{P}_{\text{S}}:~\mathop{\text{max}}_{\mathbf{w}, \bm{\gamma}} ~& \eta B\log_2(1+\gamma_{0}) + (1 - \eta) B \sum_{k=1}^K \log_2(1+\gamma_{k}) \label{obj:WSR} \\ 
\text{s.t.} ~~
& \text{SINR}_{k}^{\text{M}} \geq \gamma_{0}, ~\forall~ k \in \mathcal{K}, \label{cons:WSR-multicast-SINR}\\
& \text{SINR}_{k}^{\text{U}} \geq \gamma_{k}, ~\forall~ k \in \mathcal{K}, \label{cons:WSR-unicast-SINR}\\
& \sum_{k=0}^K \| \mathbf{w}_{k,n} \|_2^2 \leq P_n,~\forall~ n \in \mathcal{N}, \label{cons:WSR-power} \\
& \sum_{k=0}^K \left \lVert \| \mathbf{w}_{k,n} \|_2^2 \right \rVert_0 B\log_2(1+\gamma_{k}) \leq C_n,~\forall~ n \in \mathcal{N}. \label{cons:WSR-backhaul} 
\end{align} 
\end{subequations} 
Here, we have replaced the rate variables $\{r_k\}$ in $\mathcal{P}_{0}$ with the SINR variables $\{\gamma_k\}$ for convenience, where $\gamma_k = 2^{r_k} - 1$ and $\bm{\gamma} \triangleq \{\gamma_k \mid k \in \widehat{\mathcal{K}}\}$. 
Note that the peak backhaul capacity constraint \eqref{cons:WSR-backhaul} is expressed in the form of $\ell_0$-norm. Due to such backhaul constraints, each of network-wide beamforming vectors $\{\mathbf{w}_{k}\}$ may have a (group) sparse structure. 

In the following subsections, we tackle problem $\mathcal{P}_{\text{S}}$ by transforming it into a DC programming using smoothed $\ell_0$-norm approximation and some additional algebraic operations. We then obtain a stationary solution of the transformed problem by using CCP with guaranteed convergence.

\subsection{DC Transformation}
An existing approach to dealing with the non-convex SINR constraints \eqref{cons:WSR-multicast-SINR} and \eqref{cons:WSR-unicast-SINR} is to establish the equivalence between the weighted sum rate (WSR) maximization and the weighted minimum-mean-squared-error (WMMSE) minimization as in \cite{WeiYu_access,Clerckx_ICC25_joint_mult_bro_CSIT} and then apply the block coordinate descent (BCD) method.
However, due to the presence of the non-convex backhaul constraint \eqref{cons:WSR-backhaul} in our considered problem, the WMMSE-BCD method may not be applied directly.
To further deal with a similar non-convex backhaul constraint as in \eqref{cons:WSR-backhaul}, the authors in \cite{WeiYu_access} proposed to approximate the $\ell_0$-norm with a weighted $\ell_1$-norm and replace the rate function $\log_2(1+\gamma_{k})$ with the achievable rate obtained from the previous iteration. The WMMSE-BCD method is then applied \cite{WeiYu_access} (for unicast transmission only). However, no theoretical convergence is guaranteed there due to the heuristic update of the weights and rate functions \cite{Dai_Yu_ICCW15}.

In this paper, we propose to deal with all the non-convex constraints \eqref{cons:WSR-multicast-SINR}, \eqref{cons:WSR-unicast-SINR} and \eqref{cons:WSR-backhaul} by transforming them into DC forms.
Specifically, constraints \eqref{cons:WSR-multicast-SINR} and \eqref{cons:WSR-unicast-SINR} can be conveniently rewritten into DC forms as follows:
\begin{subequations}
\begin{align}
&\underbrace{\sum_{j = 1}^K \lvert \mathbf{h}_k^H \mathbf{w}_{j} \rvert^2 + \sigma_k^2}_{\text{convex}}  - \underbrace{\frac{\lvert \mathbf{h}_k^H \mathbf{w}_{0} \rvert^2} {\gamma_{0}}}_{\text{convex}} \leq 0, ~\forall~ k \in \mathcal{K},  \label{cons:WSR-DC-multicast-SINR}
\end{align}
\begin{align}
&\underbrace{\sum_{j=1,\, j \neq k}^K \lvert \mathbf{h}_k^H \mathbf{w}_{j} \rvert^2 + \sigma_k^2}_{\text{convex}}  - \underbrace{\frac{\lvert \mathbf{h}_k^H \mathbf{w}_{k} \rvert^2}{\gamma_{k}}}_{\text{convex}} \leq 0, ~\forall~ k \in \mathcal{K}. \label{cons:WSR-DC-unicast-SINR}
\end{align}
\end{subequations}
Notice that the expression $\frac{\lvert \mathbf{h}_k^H \mathbf{w}_{k} \rvert^2}{\gamma_{k}}$ is a quadratic-over-linear function, which is jointly convex in $\mathbf{w}_{k} \in \mathbb{C}^{NL\times 1}$ and $\gamma_k > 0$ \cite[Section 3.1.5]{Boyd_convex_optimization,Luong_optimal-beamforming_TSP17}. 

To transform constraint \eqref{cons:WSR-backhaul} into a DC form, we first approximate the non-smooth $\ell_0$-norm $\|x\|_0$ with a smooth, monotonically increasing, and concave function, denoted as $f_{\theta}(x)$, where $\theta > 0$ is a parameter controlling the smoothness of the approximation. In this paper, we adopt the arctangent smooth function
\begin{align}
f_{\theta}(x) = \frac{2} {\pi} \arctan \left (\frac{x}{\theta} \right),~x \geq 0, \label{cons:smooth-function}
\end{align}
which is frequently used in the literature \cite{content_centric_TWC16}. By such smooth approximation, constraint \eqref{cons:WSR-backhaul} can be approximated as
\begin{align}
\sum_{k=0}^K f_{\theta}(\| \mathbf{w}_{k,n} \|_2^2) B \log_2(1+\gamma_{k}) \leq C_n,~\forall~ n \in \mathcal{N}. \label{cons:smooth-backhaul}
\end{align}

Recall that for any $x, y$, one has $4xy = (x+y)^2 - (x-y)^2$. Therefore, by introducing two sets of auxiliary variables $\mathbf{t} \triangleq \{t_k \mid k \in \widehat{\mathcal{K}}\}$ and $\mathbf{s} \triangleq \{s_{k,n} \mid k \in \widehat{\mathcal{K}}, n \in \mathcal{N}\}$, the approximate constraint \eqref{cons:smooth-backhaul} in the product form can be equivalent to the following constraints:
\begin{subequations}
\begin{align}
& \sum_{k=0}^K B [(s_{k,n} + t_{k})^2 - (s_{k,n} - t_{k})^2] \leq 4 C_n,~\forall~ n \in \mathcal{N},  \label{cons:WSR-DC-backhaul} \\
& \log_2(1+\gamma_{k}) \leq t_{k},~\forall~ k \in \widehat{\mathcal{K}}, \label{cons:WSR-DC-rate} \\
& f_{\theta}(\| \mathbf{w}_{k,n} \|_2^2) \leq s_{k,n},~\forall~ k \in \widehat{\mathcal{K}},~n \in \mathcal{N}, \label{cons:WSR-DC-cluster} 
\end{align}
\end{subequations}
which are all in DC forms. Since $f_{\theta}$ is monotonically increasing, by introducing a set of auxiliary variables $\bm{\alpha} \triangleq \{\alpha_{k,n} \mid k \in \widehat{\mathcal{K}}, n \in \mathcal{N}\}$, constraint \eqref{cons:WSR-DC-cluster} can be further rewritten as
\begin{subequations}
\begin{align}
& f_{\theta}(\alpha_{k,n}) \leq s_{k,n},~\forall~ k \in \widehat{\mathcal{K}},~n \in \mathcal{N}, \label{cons:WSR-DC-cluster-sub1} \\
& \| \mathbf{w}_{k,n} \|_2^2 \leq \alpha_{k,n},~\forall~ k \in \widehat{\mathcal{K}},~n \in \mathcal{N}, \label{cons:WSR-DC-cluster-sub2} 
\end{align}
\end{subequations}
where constraint \eqref{cons:WSR-DC-cluster-sub2} is convex, and constraint \eqref{cons:WSR-DC-cluster-sub1} is a DC constraint. 

Finally, the original problem $\mathcal{P}_{0}$ is transformed into the following DC programming problem:
\begin{align} \label{pro:WSR-DC}
\mathcal{P}_{\text{DC}}: ~\mathop{\text{max}}_{\mathbf{w}, \bm{\gamma}, \mathbf{t}, \mathbf{s}, \bm{\alpha}} ~&  \eta B\log_2(1+\gamma_{0}) + (1 - \eta) B \sum_{k=1}^K \log_2(1+\gamma_{k}) \\
\text{s.t.} \quad~
& \eqref{cons:WSR-power}, \eqref{cons:WSR-DC-multicast-SINR}, \eqref{cons:WSR-DC-unicast-SINR}, \eqref{cons:WSR-DC-backhaul}, \eqref{cons:WSR-DC-rate}, \eqref{cons:WSR-DC-cluster-sub1}, \text{and}~\eqref{cons:WSR-DC-cluster-sub2}. \nonumber
\end{align}

\subsection{CCP Algorithm}
Problem $\mathcal{P}_{\text{DC}}$ is in the general form of DC programming where the objective function is concave and the constraints are either convex or in the DC forms, which can be efficiently solved via CCP. The main idea of CCP is to successively solve a sequence of convex subproblems, each of which is constructed by replacing the concave parts of the DC constraints with their first-order Taylor expansions \cite{yuille2003cccp}. Specifically, at the $t$-th iteration, we solve the following subproblem:
\begin{subequations} \label{pro:WSR-CCP}
\begin{align}
\mathcal{P}_{\text{DC}}^{(t)}:~\mathop{\text{max}}_{\mathbf{w}, \bm{\gamma}, \mathbf{t}, \mathbf{s}, \bm{\alpha}} ~& \eta B\log_2(1+\gamma_{0}) + (1 - \eta) B \sum_{k=1}^K \log_2(1+\gamma_{k}) \\
\text{s.t.} \quad ~
& \eqref{cons:WSR-power},~\eqref{cons:WSR-DC-cluster-sub2}, \nonumber  \\
&\sum_{j = 1}^K \lvert \mathbf{h}_k^H \mathbf{w}_{j} \rvert^2 + \sigma_k^2 - \frac{2 \Re \{ (\mathbf{w}_{0}^{(t)})^H \mathbf{h}_k \mathbf{h}_k^H \mathbf{w}_{0} \}}{\gamma_{0}^{(t)}} + \frac{\lvert \mathbf{h}_k^H \mathbf{w}_{0}^{(t)} \rvert^2 \gamma_{0}} {(\gamma_{0}^{(t)})^2} \leq 0, ~\forall~ k \in \mathcal{K}, \label{cons:WSR-CCP-multicast-SINR} \\
&\sum_{j=1,\, j \neq k}^K \lvert \mathbf{h}_k^H \mathbf{w}_{j} \rvert^2 + \sigma_k^2 - \frac{2 \Re \{ (\mathbf{w}_{k}^{(t)})^H \mathbf{h}_k \mathbf{h}_k^H \mathbf{w}_{k} \}}{\gamma_{k}^{(t)}} + \frac{\lvert \mathbf{h}_k^H \mathbf{w}_{k}^{(t)} \rvert^2 \gamma_{k}}{(\gamma_{k}^{(t)})^2} \leq 0, ~\forall~ k \in \mathcal{K}, \label{cons:WSR-CCP-unicast-SINR}\\
& \sum_{k=0}^K B [(s_{k,n} + t_{k})^2 - 2(s_{k,n}^{(t)} - t_{k}^{(t)})(s_{k,n} - t_{k}) + (s_{k,n}^{(t)} - t_{k}^{(t)})^2] \leq 4C_n,~\forall~ n \in \mathcal{N}, \label{cons:WSR-CCP-backhaul}\\
& \log_2(1+\gamma_{k}^{(t)}) + \frac{1}{(1+\gamma_{k}^{(t)})\ln2}(\gamma_{k} - \gamma_{k}^{(t)}) \leq t_{k},~\forall~ k \in \widehat{\mathcal{K}}, \label{cons:WSR-CCP-rate} \\
& f_{\theta}(\alpha_{k,n}^{(t)})  + \nabla f_{\theta}(\alpha_{k,n}^{(t)}) (\alpha_{k,n} - \alpha_{k,n}^{(t)}) \leq s_{k,n},~\forall~ k \in \widehat{\mathcal{K}}, ~n \in \mathcal{N}, \label{cons:WSR-CCP-cluster}
\end{align}
\end{subequations}
where $\{\mathbf{w}_{k,n}^{(t)}, \gamma_{k}^{(t)}, t_{k}^{(t)}, s_{k,n}^{(t)}, \alpha_{k,n}^{(t)}\}$ is the optimal solution obtained from the previous iteration. 
Problem $\mathcal{P}_{\text{DC}}^{(t)}$ is convex and can be solved using a general-purpose solver via interior-point methods \cite{Boyd_convex_optimization}. 

Note that for the general form of DC programming with DC constraints, the CCP algorithms need a feasible initial point. In our case, a feasible solution of $\mathcal{P}_{\text{DC}}$ can be obtained by generating a random initialization and scaling it to be feasible, as in \eqref{equ:BB-lower-bound-solution-r}.

We also note that due to the approximation in \eqref{cons:smooth-backhaul}, the feasible solution of problem $\mathcal{P}_{\text{DC}}$ may not be exactly feasible to the original problem $\mathcal{P}_{0}$ and hence refinement should be performed. We first determine the BS cluster for each multicast and unicast message by reserving only the links whose transmit power is larger than a certain small-value threshold, based on the solution of problem $\mathcal{P}_{\text{DC}}$. Namely, let the BS cluster be $\mathcal{S} = \{(k,n) \mid \|\mathbf{w}_{k,n}\|_2^2 \geq \epsilon^P,~k \in \widehat{\mathcal{K}}, n \in \mathcal{N}\}$, where $\epsilon^P$ is a chosen power threshold. Then we solve the following weighted sum rate maximization problem with the given BS cluster $\mathcal{S}$:
\begin{subequations} \label{pro:WSR-static}
\begin{align}
\mathcal{P}_{0}(\mathcal{S}):~\mathop{\text{max}}_{\mathbf{w}, \bm{\gamma}} ~& \eta B\log_2(1+\gamma_{0}) + (1 - \eta) B \sum_{k=1}^K \log_2(1+\gamma_{k}) \\
\text{s.t.} ~~
& \eqref{cons:WSR-multicast-SINR},~\eqref{cons:WSR-unicast-SINR},~\eqref{cons:WSR-power}, \nonumber \\
& \sum_{(k,n) \in \mathcal{S}} B \log_2(1+\gamma_{k}) \leq C_n,~\forall~ n \in \mathcal{N}, \label{cons:WSR-static-backhaul}  \\
& \mathbf{w}_{k,n}  = \mathbf{0}_L,~\forall~ (k,n) \notin \mathcal{S},
\end{align}
\end{subequations}
which is a DC programming problem and can be directly solved via CCP. Empirically, the algorithm would converge within just a few iterations if the initial point is constructed from the solution of problem $\mathcal{P}_{\text{DC}}$.

The overall CCP-based algorithm for solving problem $\mathcal{P}_{0}$ is summarized in Alg. \ref{alg:WSR-DC-CCP}.
\begin{algorithm}[ht]
\caption{The CCP-based algorithm for solving problem $\mathcal{P}_{0}$} \label{alg:WSR-DC-CCP}
\begin{algorithmic}[0]
\STATE \textbf{Initialization:} Randomly generate an initial point and scale it to be feasible, denoted as $\{\mathbf{w}^{(0)}, \bm{\gamma}^{(0)}, \mathbf{t}^{(0)}, \mathbf{s}^{(0)}, \bm{\alpha}^{(0)}\}$. Set the power threshold $\epsilon^P$ and the iteration index $t \leftarrow 0$.
\STATE \textbf{Repeat}
\begin{enumerate}
	\item Update $\{\mathbf{w}^{(t+1)}, \bm{\gamma}^{(t+1)}, \mathbf{t}^{(t+1)}, \mathbf{s}^{(t+1)}, \bm{\alpha}^{(t+1)}\}$ via solving problem $\mathcal{P}_{\text{DC}}^{(t)}$.
	\item Set $t \leftarrow t + 1$.
\end{enumerate}
\STATE \textbf{Until} the stopping criterion is met. Denote the solution as $\widehat{\mathbf{w}}$.
\STATE \textbf{Refinement}
\begin{enumerate}
	\item Determine the BS cluster $\mathcal{S}$ based on $\widehat{\mathbf{w}}$.
	\item Solve problem $\mathcal{P}_{0}(\mathcal{S})$ via CCP.
\end{enumerate}
\end{algorithmic}
\end{algorithm}

\subsection{Convergence and Complexity Analysis}
\subsubsection{Convergence}
With a feasible initial point, the CCP iteration is guaranteed to converge to a stationary solution of problem $\mathcal{P}_{\text{DC}}$. Note that the obtained stationary solution of problem $\mathcal{P}_{\text{DC}}$ is not necessarily a stationary solution of the original problem $\mathcal{P}_{\text{S}}$ due to the approximation in \eqref{cons:smooth-backhaul}. Intuitively, we can see that at each iteration of CCP, the optimal solution obtained from the previous iteration, i.e., $\{\mathbf{w}^{(t)}, \bm{\gamma}^{(t)}, \mathbf{t}^{(t)}, \mathbf{s}^{(t)}, \bm{\alpha}^{(t)}\}$, is a feasible solution of the subproblem $\mathcal{P}_{\text{DC}}^{(t)}$. The achieved objective at the current iteration should be not smaller than the one at the previous iteration. Therefore, the objective value is non-deceasing and will converge. We refer the interested reader to \cite{lanckriet2009convergence} for a rigorous proof of the convergence.

\subsubsection{Complexity Analysis}
At each CCP iteration, we need to solve a convex subproblem $\mathcal{P}_{\text{DC}}^{(t)}$. With log functions in the objective, $\mathcal{P}_{\text{DC}}^{(t)}$ can be approximated by a sequence of SOCPs \cite{Boyd_SOCP_1998} via the successive approximation method \cite{cvx}. Each SOCP can then be solved with a complexity of $\mathcal{O}((KNL)^{3.5})$ via a general-purpose solver, e.g., SDPT3 in CVX \cite{cvx} as we use in the simulation part of this paper.
Suppose that the CCP requires $T_{\text{max}}^C$ iterations to converge, the worst-case computational complexity is therefore $\mathcal{O}(T_{\text{max}}^C (KNL)^{3.5})$. 
Compared with the proposed BB-based algorithm in Section \ref{sec:BB-based Optimal Algorithm}, the proposed CCP-based algorithm can converge much faster in practice, i.e., $T_{\text{max}}^C \ll T_{\text{max}}^B$, which is more efficient in terms of complexity. 

\section{Simulation Results} \label{sec:Simulation Results}
In this section, numerical simulations are provided to verify the effectiveness of the proposed algorithms. The superiority of the proposed non-orthogonal multicast and unicast transmission scheme over the orthogonal transmission schemes is also demonstrated. 
We consider a hexagonal multi-cell cellular network, denoted as $(N, K, L)$, where there are $N$ BSs and $K$ mobile users, and each BS is equipped with $L$ antennas and located at the center of the cell. The distance between adjacent BSs is set to $500$ m. The mobile users are uniformly and randomly distributed in the network, excluding an inner circle of $50$ m around each BS. 
The transmit antenna power gain is $9$ dBi. The available bandwidth of the wireless channel is $B = 10$ MHz. The small-scale fading is generated from the normalized Rayleigh fading. The path loss is modeled as $148.1 + 37.6 \log_{10}(d)$ in dB, where $d$ is the distance in km. The standard deviation of log-normal shadowing is $8$ dB. The noise power spectral density $\sigma_k^2$ is $-174$ dBm/Hz for all users. 
The weighting parameter between the multicast rate and the unicast rate is set to $\eta = 0.9$ if not specified otherwise.
The tolerance of the gap between the upper bound and the lower bound in Alg. \ref{alg:WSR-WSR-Clustering-BB} is set as $\epsilon = 10^{-2}$.
The power threshold for refinement in Alg. \ref{alg:WSR-DC-CCP} is set as $\epsilon^P = -30$ dBm. The CCP iteration stops when the relative increase of the objective value is less than $10^{-3}$ or when a maximum of $40$ iterations is reached. The smoothness parameter in \eqref{cons:smooth-function} is set as $\theta = 10^{-6}$. For simplicity, all BSs have the same maximum transmit power and the same maximum backhaul capacity, i.e., $P_n = P$ and $C_n = C$, for all $n \in \mathcal{N}$.
The plots in Section \ref{sec:Convergence-Behavior-of-the-Proposed-Algorithms} are based on a random channel realization. The plots in Sections \ref{sec:Effectiveness-of-the-Proposed-Algorithms} and \ref{sec:Performance-Comparison-with-the-Orthogonal-Scheme} are obtained by averaging over $100$ independent channel realizations.

\subsection{Convergence Behavior of the Proposed Algorithms} \label{sec:Convergence-Behavior-of-the-Proposed-Algorithms}
\begin{figure}[tbp]
\begin{centering}
\includegraphics[scale=.36]{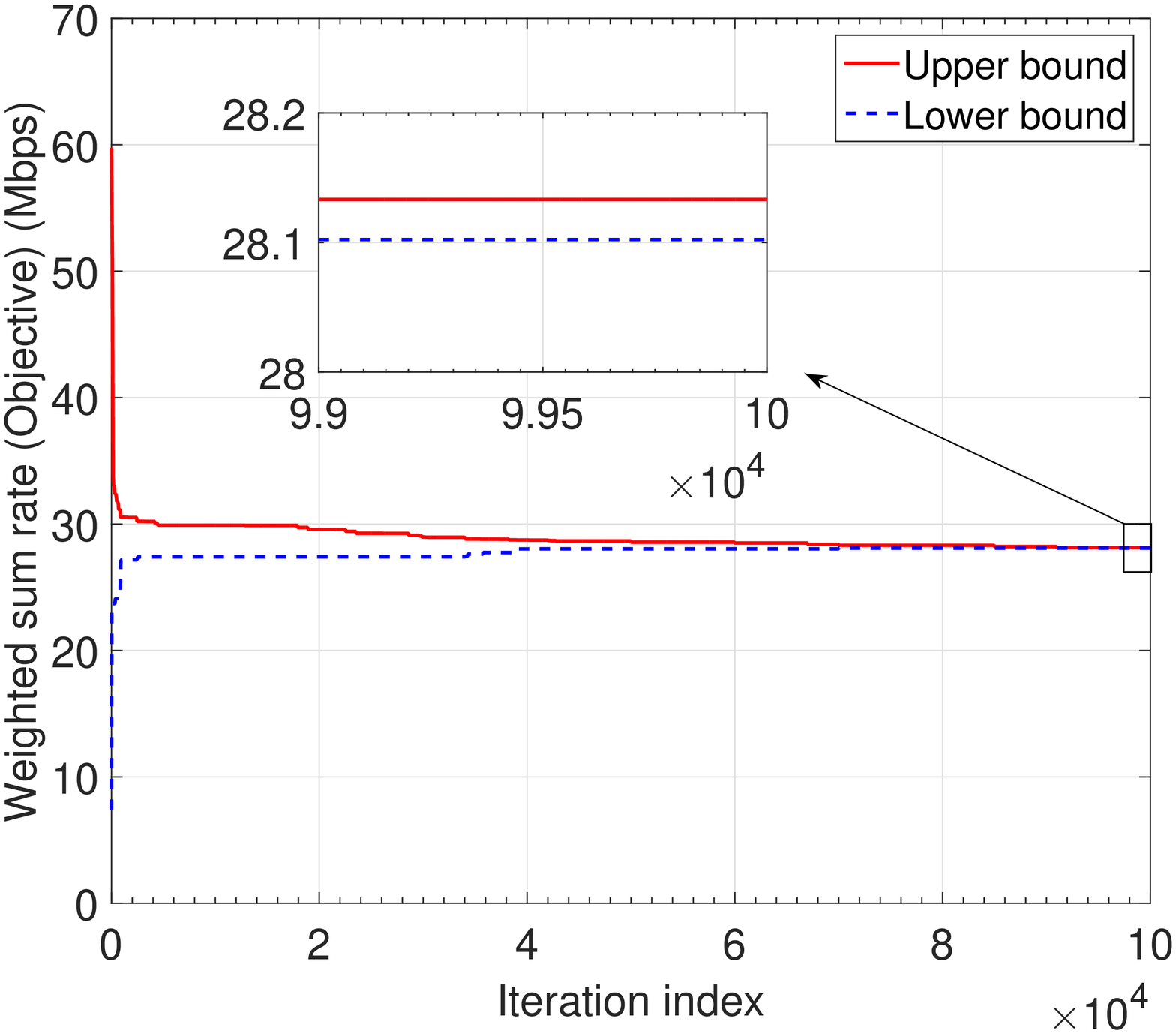}
 \caption{\small{Convergence behavior of the proposed BB-based algorithm.}}\label{fig:fig_convergence_BB}
\end{centering}
\end{figure}
In this subsection, we demonstrate the convergence behaviors of the proposed BB-based and CCP-based algorithms. We generate a problem instance in a small-scale network with $(N,K,L) = (3,2,2)$ and solve it using both the proposed BB-based and CCP-based algorithms. The maximum transmit power of each BS is set as $P = 20$ dBm and the backhaul capacity is $C = 100$ Mbps. Fig. \ref{fig:fig_convergence_BB} shows the upper bounds $\{\Phi_{\text{U}}^{t}\}$ and the lower bounds $\{\Phi_{\text{L}}^{t}\}$ of the weighted sum of the multicast rate and the unicast rate (i.e. the objective) in the BB-based algorithm. 
We can see that the upper bound and the lower bound are non-increasing and non-decreasing, respectively, and the gap between them is reduced rapidly during first few iterations since a large number of infeasible subregions are removed. This gap becomes smaller as the iteration index increases. Although the number of iterations $T_{\text{max}}^B$ in \eqref{equ:maximum_iteration} can be very large if the tolerance $\epsilon$ is small, which is not practical due to the prohibitively high complexity, the achieved results can still be used as the network performance benchmark. 

\begin{figure}[tbp]
\begin{centering}
\includegraphics[scale=.36]{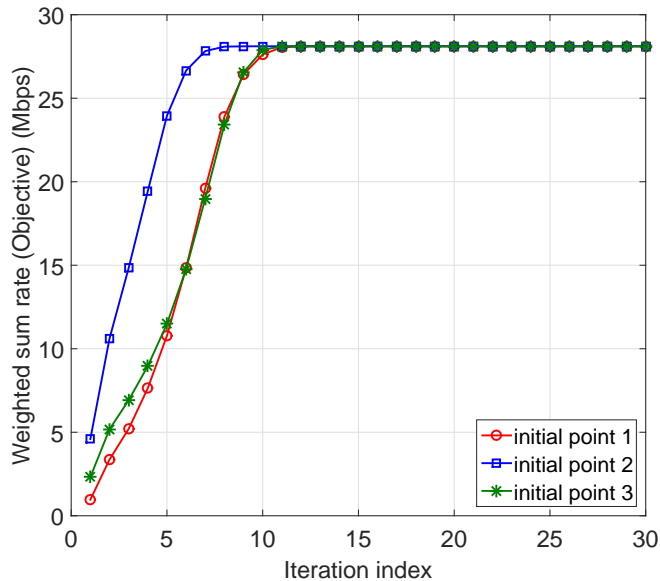}
 \caption{\small{Convergence behavior of the proposed CCP-based algorithm.}}\label{fig:fig_convergence_CCP}
\end{centering}
\end{figure}
Fig. \ref{fig:fig_convergence_CCP} shows the weighted sum of the multicast rate and the unicast rate achieved by the CCP-based algorithm with three different initial points. We observe that with different initial points, the objectives of the proposed CCP-based algorithm converge to a same value, all within $15$ iterations.
Compared with the proposed BB-based algorithm, the proposed CCP-based algorithm can converge much faster, i.e., $T_{\text{max}}^C \ll T_{\text{max}}^B$, which is more efficient in terms of complexity. 

\subsection{Effectiveness of the Proposed Algorithms} \label{sec:Effectiveness-of-the-Proposed-Algorithms}
\begin{figure}[tbp]
\begin{centering}
\includegraphics[scale=.36]{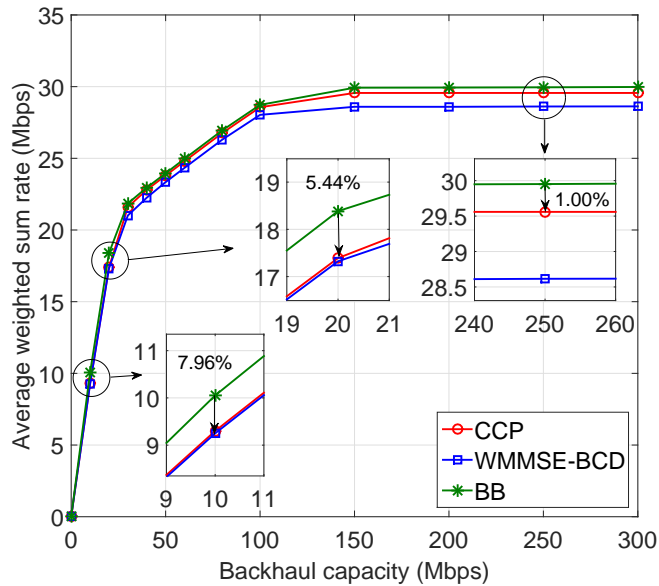}
 \caption{\small{Average weighted sum rate versus backhaul capacity $C$ for different benchmarks with $(N,K,L) = (3,2,2)$.}}\label{fig:fig_cluster_smallscale}
\end{centering}
\end{figure}
We first demonstrate the performances of the proposed BB-based and CCP-based algorithms in a small-scale network with $(N,K,L) = (3,2,2)$ in Fig. \ref{fig:fig_cluster_smallscale}.
As a benchmark, we consider generalizing the WMMSE-BCD algorithm in \cite{WeiYu_access} to solve problem $\mathcal{P}_{0}$. Note that the convergence of this algorithm has not been established theoretically, thus we set the maximum number of iterations to $40$.
Fig. \ref{fig:fig_cluster_smallscale} illustrates the average weighted sum of the multicast rate and the unicast rate with different backhaul capacities $C$. The maximum transmit power of each BS is set as $P = 20 $ dBm.
We first observe that the CCP-based achieves high performance that is close to the optimal BB-algorithm when the backhaul constraint is not stringent (e.g., $C \geq 30$ Mbps). For example, there is only $1.00 \%$ performance loss when $C = 250$ Mbps. Our proposed CCP-based algorithm is better than the WMMSE-BCD algorithm, especially for large backhaul capacities (e.g., $C \geq 150$ Mbps). 
We also observe that when the backhaul constraint is stringent (e.g., $C \leq 20$ Mbps), there is a relatively large performance gap to the optimal BB-based algorithm for both of the CCP-based and WMMSE-BCD algorithms. This might be because that when the backhaul constraint is stringent, there is little room for rate maximization and both CCP-based and WMMSE-BCD algorithms are more likely to get stuck in unfavorable local solutions.

\begin{figure}[tpb]
\begin{centering}
\includegraphics[scale=.36]{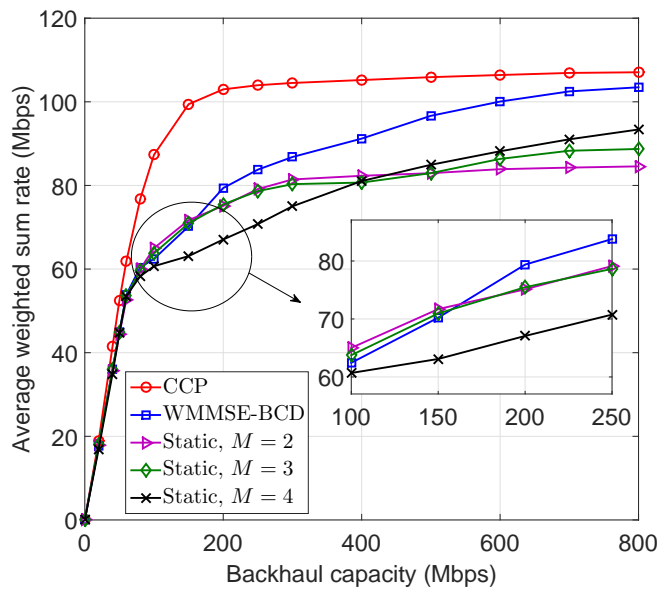}
 \caption{\small{Average weighted sum rate versus backhaul capacity $C$ for different benchmarks with $(N,K,L) = (7,10,4)$.}}\label{fig:fig_cluster}
\end{centering}
\end{figure}

We then demonstrate the performances of the proposed CCP-based algorithm in a larger network with $(N,K,L) = (7,10,4)$ in Fig. \ref{fig:fig_cluster}. The BB-based algorithm is not considered due to its high complexity. Besides WMMSE-BCD, we also consider a benchmark algorithm with static BS clustering $\mathcal{S}$, where the multicast message is transmitted by all BSs via full cooperation and each unicast message is transmitted by a static cluster of BSs that are closest to the user with the cluster size $M \in \{2,3,4\}$ via partial cooperation. The maximum transmit power of each BS is set as $P = 30 $ dBm. 
From Fig. \ref{fig:fig_cluster}, it can be seen that all the considered algorithms do not differ much when the backhaul capacity is small (e.g., $C < 50$ Mbps). This is expected because there is little room for rate maximization when the backhaul constraint is stringent. While when the backhaul capacity is large (e.g., $C > 100$ Mbps), the proposed algorithm is superior to all the benchmarks, especially the static BS clustering schemes. This clearly demonstrates the effectiveness of our proposed algorithm.
It can also be seen that among all the considered static BS clustering schemes, the best cluster size $M$ varies at different backhaul constraints. This is due to the well-known tradeoff that allowing more BSs for joint transmission increases the transmission rate but at the expense of higher backhaul consumption.

\begin{table}[tbp]
\caption{Average cluster size obtained by the CCP-based algorithm.}  \label{tab:average-cluster-size}
\centering
\begin{tabular}{|c|c|c|c|c|c|c|c|c|c|c|c|c|c|}
\hline
Backhaul (Mbps)& 50 & 100 & 150 & 200 & 250 & 300 & 400 & 500 & 600 & 700 & 800 \\
\hline
Multicast & 5.10 & 6.83 & 7 & 7 & 7 & 7 & 7 & 7 & 7 & 7 & 7 \\
\hline
Unicast & 0.45 & 0.84 & 2.06 & 3.14 & 4.04 & 4.69 & 5.65 & 6.15 & 6.67 & 6.96 & 6.99  \\
\hline
\end{tabular}
\end{table}

From Fig. \ref{fig:fig_cluster}, we also observe that at small backhaul capacity region (e.g., $C < 100$ Mbps), the average weighted sum rate of the CCP-based algorithm increases almost linearly when $C$ increases. This suggests that the system is backhaul limited. However, at large backhaul capacity region (e.g., $C > 200$ Mbps), the average weighted sum rate approaches constant when $C$ further increases. This means that the system becomes power limited.

Finally, we report in Table \ref{tab:average-cluster-size} the average cluster size (i.e., the number of serving BSs) of the multicast message and the average per-user cluster size of the unicast messages obtained by the CCP-based algorithm. Note that when the backhaul constraint is extremely stringent (i.e., $< 100$ Mbps), some users cannot be served for unicast in the sum rate maximization problem, and accordingly, the actual cluster sizes of those users are $0$. Thus, the average per-user cluster size of the unicast service can be less than one. 
From Table \ref{tab:average-cluster-size}, it can be seen that the cluster size of the multicast message is always larger than $5$ for different backhaul capacities. This is largely because the users are uniformly and randomly distributed in the network and most of the BSs should be involved to efficiently deliver the multicast message. For the unicast message, the cluster size increases with the backhaul capacity as expected. When the backhaul capacity is sufficiently large (i.e., $>700$ Mbps), all the BSs participate in delivering the unicast message.

\subsection{Performance Comparison with the Orthogonal Scheme} \label{sec:Performance-Comparison-with-the-Orthogonal-Scheme}
\begin{figure}[tbp]
\begin{centering}
\includegraphics[scale=.36]{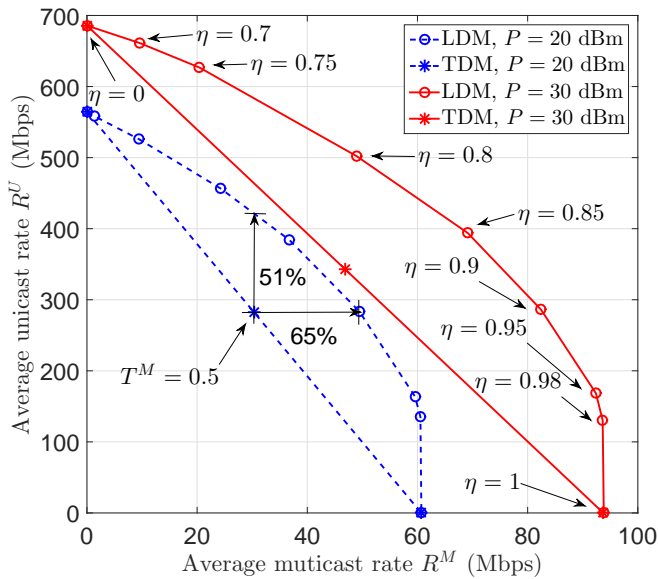}
 \caption{\small{Comparison of the average multicast and unicast rate region between LDM and TDM.}} \label{fig:fig_tradeoff}
\end{centering}
\end{figure}
The comparison of the achievable multicast-unicast rate region between the LDM-based non-orthogonal scheme and the TDM-based orthogonal scheme is illustrated in Fig. \ref{fig:fig_tradeoff} with $(N,K,L) = (7,10,4)$.
Recall that the achievable multicast and unicast rates are defined as $R^{\text{M}} \triangleq B r_{0}$ and $R^{\text{U}} \triangleq B \sum_{k=1}^K r_{k}$, respectively. The backhaul capacity is set as $C = 200$ Mbps. For LDM, the multicast-unicast tradeoff curves are obtained by controlling the weighting parameter $\eta$ in $\mathcal{P}_{0}$. When $\eta = 1$ or $0$, the objective only accounts for the multicast rate or the unicast rate.
For TDM, let $T^M \in [0,1]$ denote the fraction of time devoted to the multicast transmission.
Note that $\eta = 1$ and $\eta = 0$ for LDM are equivalent to $T^M = 1$ and $T^M = 0$ for TDM, respectively. It is obvious that the achievable multicast-unicast rate region of LDM is much larger than that of TDM. More specifically, when $P = 20$ dBm and $T^M = 0.5$, compared with TDM, LDM increases the unicast rate from $282$ Mbps to $425$ Mbps with the same multicast rate of $30.4$ Mbps, which is $51\%$ higher.
And it also increases the multicast rate from $30.4$ Mbps to $50.0$ Mbps with the same unicast rate of $282$ Mbps, resulting in $65\%$ improvement.


\section{Conclusion} \label{sec:Conclusion}
This paper proposed to incorporate multicast and unicast services into cellular networks using a non-orthogonal transmission scheme based on the LDM principle. We formulated an optimization problem to maximize the weighted sum of the multicast rate and the unicast rate subject to the peak power constraint and the peak backhaul constraint for each BS via joint BS clustering and beamforming design. The formulated non-convex MINLP problem is optimally solved using the proposed BB-based algorithm. We also proposed a low-complexity algorithm to find a high-performance solution by means of sparse optimization and CCP. Simulation results demonstrated that our proposed LDM-based non-orthogonal scheme can significantly outperform orthogonal schemes in terms of the achievable multicast-unicast rate region.

\appendices
\section{Proof of Lemma \ref{lemma_convergence_BB}} \label{proof_of_lemma_convergence_BB}
To prove Lemma \ref{lemma_convergence_BB}, we first provide the following lemma:
\begin{lemma} \label{lemma_multicast_relaxation}
Given any rate $\underline{r}_{0} > 0$ and interval $[\underline{\varphi}_{k}, \bar{\varphi}_{k}]$ with $\bar{\varphi}_{k} - \underline{\varphi}_{k} \leq \pi$, for all $\mathbf{w} \in Conv(\mathcal{D}_{k}^{[\underline{\varphi}_{k}, \bar{\varphi}_{k}]}(\underline{r}_{0}))$ in \eqref{cons:WSR-Clustering-multicast-SINR-SOC-envelope}, we have
\begin{align} 
\frac {\lvert \mathbf{h}_k^H \mathbf{w}_{0} \rvert^2} {\sum_{i = 1}^K \lvert \mathbf{h}_k^H \mathbf{w}_{i} \rvert^2 + \sigma_k^2} \geq (2^{\underline{r}_{0}}-1) \cos^2 \left( \frac{\bar{\varphi}_{k}-\underline{\varphi}_{k}}{2} \right).
\end{align}
\end{lemma}

This lemma is an extension of \cite[Proposition 2]{Liu_optimalmulticast_TSP17}. We omit its proof here for brevity.
For any $\mathcal{Q} \subseteq \mathcal{Q}_{\text{init}}$, assume $\mathcal{Q} = [\mathbf{a}, \mathbf{b}]$, where $\mathbf{a} = [\underline{\mathbf{s}}^T, \underline{\mathbf{r}}^T, \underline{\bm{\varphi}}^T]^T$ and $\mathbf{b} = [\bar{\mathbf{s}}^T, \bar{\mathbf{r}}^T, \bar{\bm{\varphi}}^T]^T$. Since $\text{size}(\mathcal{Q}) \leq \delta$, we have $\max_{1\leq j\leq N_q} \{b_j - a_j\} \leq \delta$. 
Next, based on the above result, we first estimate the upper bound $\Phi_{\text{U}}(\mathcal{Q})$ and the lower bound $\Phi_{\text{L}}(\mathcal{Q})$ and then estimate its gap.

{\em Upper Bound:} Let $\{\mathbf{w}^*, \mathbf{r}^*, \mathbf{s}^*\}$ be the optimal solution of problem \eqref{pro:WSR-Clustering-Upper} \footnote{We ignore the case where problem \eqref{pro:WSR-Clustering-Upper} is infeasible, since in this case the box $\mathcal{Q}$ does not contain the optimal solution.}, the upper bound obtained by solving problem \eqref{pro:WSR-Clustering-Upper} is given by $\Phi_{\text{U}}(\mathcal{Q}) = \eta K B r_{0}^* + (1 - \eta) B\sum_{k=1}^K r_{k}^*$, which satisfies $\Phi_{\text{U}}(\mathcal{Q}) \leq \eta K B \bar{r}_{0} + (1 - \eta) B \sum_{k=1}^K \bar{r}_{k}$ according to constraint \eqref{cons:WSR-Clustering-binary-relaxed}. 

{\em Lower Bound:} Since the size of the box $\mathcal{Q}$ is small enough, i.e., $\delta < 1$, we have $\underline{s}_{k,n} = \bar{s}_{k,n}$ by the splitting rule \eqref{cons:WSR-Clustering-splitting-rule}. Therefore, the optimal solution $s_{k,n}^*$ of problem \eqref{pro:WSR-Clustering-Upper} is $s_{k,n}^* = \underline{s}_{k,n} = \bar{s}_{k,n} = 0 \text{ or } 1$. In \eqref{equ:BB-lower-bound-solution}, let $p_j$ be the smallest non-zero element of $\{\|\mathbf{w}_{k,n}^*\|_2^2\}$, we have $\tilde{s}_{k,n} = s_{k,n}^*$ and $\tilde{\mathbf{w}}_{k,n} = \mathbf{w}_{k,n}^*$. Moreover, according to Lemma \ref{lemma_multicast_relaxation}, we have
\begin{align} 
\frac {\lvert \mathbf{h}_k^H \mathbf{w}_{0}^* \rvert^2} {\sum_{i = 1}^K \lvert \mathbf{h}_k^H \mathbf{w}_{i}^* \rvert^2 + \sigma_k^2} \geq (2^{\underline{r}_{0}}-1) \cos^2 \left( \frac{\bar{\varphi}_{k}-\underline{\varphi}_{k}}{2} \right).
\end{align}
Then, the multicast rate in \eqref{equ:BB-lower-bound-solution-R0} satisfies
\begin{subequations}
\begin{align} 
r_{0}(\tilde{\mathbf{w}}) = r_{0}(\mathbf{w}^*) &= \min_{k \in \mathcal{K}} \log_2 \left(1 + \frac {\lvert \mathbf{h}_k^H \tilde{\mathbf{w}}_{0} \rvert^2} {\sum_{i = 1}^K \lvert \mathbf{h}_k^H \tilde{\mathbf{w}}_{i} \rvert^2 + \sigma_k^2} \right) \\
&\geq \min_{k \in \mathcal{K}} \log_2 \left(1 + (2^{\underline{r}_{0}}-1) \cos^2 \left( \frac{\bar{\varphi}_{k}-\underline{\varphi}_{k}}{2} \right) \right) \\
&= \log_2 \left(1 + (2^{\underline{r}_{0}}-1) \cos^2 \left( \frac{\max_{k \in \mathcal{K}}(\bar{\varphi}_{k}-\underline{\varphi}_{k})}{2} \right) \right) \\
&\geq \log_2 \left(2^{\underline{r}_{0}} \cos^2 \left( \frac{\max_{k \in \mathcal{K}}(\bar{\varphi}_{k}-\underline{\varphi}_{k})}{2} \right) \right) \\
&= \underline{r}_{0} + 2\log_2 \left(\cos \left( \frac{\max_{k \in \mathcal{K}}(\bar{\varphi}_{k}-\underline{\varphi}_{k})}{2} \right) \right).
\end{align}
\end{subequations}
Similarly, the unicast rate in \eqref{equ:BB-lower-bound-solution-Rk} satisfies $r_{k}(\tilde{\mathbf{w}}) = r_{k}(\mathbf{w}^*) \geq \underline{r}_{k}$.
Let $\tilde{r}_{0} = \underline{r}_{0} + 2\log_2 \left(\cos \left( \frac{\max_{k \in \mathcal{K}}(\bar{\varphi}_{k}-\underline{\varphi}_{k})}{2} \right) \right)$ and $\tilde{r}_{k} = \underline{r}_{k}$. Since $\log_2 \left(\cos \left( \frac{\max_{k \in \mathcal{K}}(\bar{\varphi}_{k}-\underline{\varphi}_{k})}{2} \right) \right) < 0$, there is $\tilde{r}_{0} \leq \underline{r}_{0}$. Thus, the backhaul constraint \eqref{cons:WSR-Clustering-backhaul} is satisfied, i.e., $\sum_{k=0}^K \tilde{s}_{k,n} \tilde{r}_{k} \leq \sum_{k=0}^K \tilde{s}_{k,n} \underline{r}_{k}  \leq C_n$ for all $n \in \mathcal{N}$. Then, $\{\tilde{\mathbf{w}}, \tilde{\mathbf{s}}, \tilde{\mathbf{r}}\}$ is a feasible solution of the original problem $\mathcal{P}_{0}$. The lower bound is given by $\Phi_{\text{L}}(\mathcal{Q}) = \eta K B\tilde{r}_{0} + (1 - \eta) B\sum_{k=1}^K \tilde{r}_{k}$. 

Finally, the gap between the upper bound $\Phi_{\text{U}}(\mathcal{Q})$ and the lower bound $\Phi_{\text{L}}(\mathcal{Q})$ is given by
\begin{subequations}
\begin{align} 
\Phi_{\text{U}}(\mathcal{Q}) - \Phi_{\text{L}}(\mathcal{Q}) &=\eta K B(r_{0}^* - \tilde{r}_{0}) + (1 - \eta) B\sum_{k=1}^K (r_{k}^* - \tilde{r}_{k})\\
&\leq \eta K B(\bar{r}_{0} - \underline{r}_{0}) + (1 - \eta) B\sum_{k=1}^K (\bar{r}_{k} - \underline{r}_{k}) \nonumber \\
& \quad - 2 \eta KB\log_2 \left(\cos \left( \frac{\max_{k \in \mathcal{K}}(\bar{\varphi}_{k}-\underline{\varphi}_{k})}{2} \right) \right)\\
&\leq K B\delta - 2\eta KB\log_2 \left(\cos(\delta/2)\right).
\end{align}
\end{subequations}
Since the function $g(\delta) = KB\delta -2\eta KB\log_2 \left(\cos(\delta/2) \right)$ is monotonically increasing for all $\delta \in (0,1)$, there always exists a small enough $\delta$ such that $g(\delta) \leq \epsilon$, which can be found by bisection search.

\section{Proof of Lemma \ref{lemma_complexity_BB}} \label{proof_of_lemma_complexity_BB}
We prove Lemma \ref{lemma_complexity_BB} based on the contradiction principle. Suppose that Alg. \ref{alg:WSR-WSR-Clustering-BB} does not terminate within $T_{\text{max}}^B$ iterations. Then, according to Lemma \ref{lemma_convergence_BB}, we conclude that the selected box at the $t$-th iteration satisfies $\text{size}(\mathcal{Q}^{*}) > \delta$ for all $t = 1, 2, \dots, T_{\text{max}}^B$.  
If the longest edge chosen to be split satisfies $j^{*} > (K+1)N$, then, after the splitting, the width of the $j^{*}$-th edge of the two boxes $\mathcal{Q}^{*}_{(1)}$ and $\mathcal{Q}^{*}_{(2)}$ is greater than $\delta/2$. Similarly, for each box $\mathcal{Q}$ partitioned from the original box $\mathcal{Q}_{\text{init}}$, there holds $b_j - a_j > \delta/2$ for all $j > (K+1)N$. Hence, the volume of each box $\mathcal{Q}$ is not less than $(\frac{\delta}{2})^{(K-1) + (K+1)}$. Note that due to the binary nature of the variable $\mathbf{s}$, the volume of a box $\mathcal{Q}$ is calculated without taking the variable $\mathbf{s}$ into account. If the longest edge $j^{*} \leq (K+1)N$, we get two boxes with the same volume after the splitting. At the $T_{\text{max}}^B$-th iteration, the total volume of all $T_{\text{max}}^B$ boxes is not less than $T_{\text{max}}^B (\frac{\delta}{2})^{(K-1) + (K+1)}$. Obviously, the volume of $\mathcal{Q}_{\text{init}}$ is $2^{(K+1)N} (2\pi)^{K-1} \prod_{k=0}^{K} r_{\text{max}}^k$. By the choice of $T_{\text{max}}^B$, we get $T_{\text{max}}^B (\frac{\delta}{2})^{(K-1) + (K+1)} > 2^{(K+1)N} (2\pi)^{K-1} \prod_{k=0}^{K} r_{\text{max}}^k$, which implies that the total volume of all $T_{\text{max}}^B$ boxes is greater than that of the original box $\mathcal{Q}_{\text{init}}$. This is a contradiction. Hence, the algorithm will terminate within at most $T_{\text{max}}^B$ iterations.

\bibliographystyle{IEEEtran}
\bibliography{IEEEabrv,non-orth_multicast_unicast}

\end{document}